\documentclass[superscriptaddress,nobibnotes,prd,nofootinbib]{revtex4}
\usepackage{amsmath}

\DeclareMathOperator{\sech}{sech}
\usepackage{amsfonts}
\usepackage{amssymb}
\usepackage{graphicx}
\makeatletter
\newcommand*{\rom}[1]{\expandafter\@slowromancap\romannumeral #1@}
\makeatother

\usepackage{dcolumn}
\usepackage{bm}
\usepackage{color}
\usepackage[normalem]{ulem}
\usepackage[percent]{overpic}
\usepackage{natbib}
\usepackage[usenames,dvipsnames]{xcolor}
\usepackage{hyperref}
\hypersetup{colorlinks=true,
citecolor=MidnightBlue, linkcolor=MidnightBlue, urlcolor=MidnightBlue}
\usepackage{tikz}
\definecolor{purple}{rgb}{1,0,1}
\definecolor{lime}{HTML}{A6CE39} 

\usepackage{epsfig}

\newcommand{\orcidicon}{%
	\begin{tikzpicture}
	\draw[lime, fill=lime] (0,0)
		circle [radius=0.16]
		node[white] {{\fontfamily{qag}\selectfont \tiny ID}};
	\draw[white, fill=white] (-0.0625,0.095)
		circle [radius=0.007];
	\end{tikzpicture}	\hspace{-2mm}
}

\newcommand\orcidMarziyeh{{\href{https://orcid.org/0000-0003-2736-9396}{\orcidicon}}}
\newcommand\orcidFrancisco{{\href{https://orcid.org/0000-0002-9388-8373}{\orcidicon}}}

\begin{document}

\title{Thick branes via higher order field theory models with exponential and power-law tails}

\author{Marzieh Peyravi\orcidMarziyeh\!\!}
\email{marziyeh.peyravi@stu-mail.um.ac.ir}
\affiliation{Department of Physics, School of Sciences, Ferdowsi University
of Mashhad, Mashhad 91775-1436, Iran}

\author{Samira Nazifkar}
\email{nazifkar@neyshabur.ac.ir}
\affiliation{Faculty of Science, Department of Physics, University of Neyshabur, Neyshabur, Iran}

\author{Francisco S. N. Lobo\orcidFrancisco\!\!}
\email{fslobo@fc.ul.pt}
\affiliation{Instituto de Astrof\'{\i}sica e Ci\^{e}ncias do Espa\c{c}o, Faculdade de
Ci\^encias da Universidade de Lisboa, Campo Grande, Edif\'{\i}cio C8,
P-1749-016 Lisbon, Portugal}
\affiliation{Departamento de F\'{i}sica, Faculdade de Ci\^{e}ncias da Universidade de Lisboa, Campo Grande, Edif\'{\i}cio C8, P-1749-016 Lisbon, Portugal}

\author{Kurosh Javidan}
\email{javidan@um.ac.ir}
\affiliation{Department of Physics, School of Sciences, Ferdowsi University
of Mashhad, Mashhad 91775-1436, Iran}


\begin{abstract}
In this work, we obtain exact thick brane models in $4+1$ dimensions generated by higher order field theory kinks, inspired by specific potentials for $\phi^{10}$ and $\phi^{18}$ models. 
We verify that the geodesic equation along the fifth dimension confirms the confining effects of the scalar field on the brane for all of these models. These models provide new solutions with exponential and power-law tails which live in different topological sectors. We show that the resulting branes of specific exponential law models do not possess $Z_2$-symmetry. Furthermore, we examine the stability of the thick branes, by determining the sign of the $w^2$ term in the expansion of the potential for the resulting Schr\"{o}dinger-like equation. It turns out that two of the three models of the $\phi^{10}$ brane are stable, while another contains unstable modes for certain ranges of the model parameters. We also show that the brane solution from the specific $\phi^{18}$ models are stable, while the others involve neutral equilibrium. The asymptotic behaviour of the brane solutions are also discussed.

\medskip


\end{abstract}
\date{\today }
\maketitle


\section{Introduction}\label{Intr}

The braneworld scenario has been built on the concept of a 4-dimensional hyper-surface observable universe embedded in a higher-dimensional space denoted the ``bulk'', in such a way that all particles and fields are trapped on the brane except gravity \cite{Randall:1999ee, Randall:1999vf, Dvali:2000hr, Shiromizu:1999wj, Maartens:2003tw,Ge2, lin,PCU, Ge3}. In this context, a plausible extension to the standard model of particle physics is the existence of extra dimensions \cite{Ge1,Ge5}. Most of brane models contain one or more discontinuous 3-branes in the bulk \cite{de}, where two 3-branes models can describe the large hierarchy between the scales of weak and gravitational forces \cite{de,CG}. It is widely assumed that in braneworld scenarios $Z_{2}$-symmetry is imposed \cite{blg,AIS,DM,CC} and under this symmetry, which originated in $M$-theory, the bulk metric on both sides of the brane should be the same \cite{AIS,CC}. Moreover, under such a symmetry the empty bulk on either sides of the brane have the same negative cosmological constant and as a result, they are AdS \cite{DM}.
There are, however, brane models in which there is no $Z_{2}$-symmetry and the bulk is different on both sides of the brane \cite{AIS}, where the cosmological constant may differ on the two sides \cite{JL}, and the possibility that the bulk may have varying cosmological constant has also been analysed \cite{CC}.

Since topological defects, such as scalar field solitons are effective and stable structures to generate branes, it seems that extra-dimensional models, in this context, require the existence of scalar fields \cite{de, lin,PCU}. On the other hand, scalar fields also serve to stabilize the size of the compact extra dimension \cite{lin,PCU} and help modify the Randall-Sundrum warped-space \cite{Ge3} to a smoothed-out version \cite{de}, or to cut off the
extra dimension at a singularity \cite{CG,Ge1}. As one can show, the non-linearity in the scalar field and in particular the existence of discrete vacua in the self-interaction of the scalar field lead to the appearance of a stable localized solution, which motivates the construction of thick brane models \cite{blg}. In this regard, a general method for determining the lowest energy configuration has been analysed in \cite{Ge4,br}. It is interesting to note that thick branes have also been considered in modified theories of gravity \cite{Rosa:2021tei,Bazeia:2020jma,Rosa:2020uli,Rosa:2022fhl,Bazeia:2022agk}.

Replacing solitons as branes have been studied in many references, where some of these models consider polynomial kinks such as the $\phi^{4}$ and $\phi^{6}$ models \cite{Dzh,Bazeia:2004dh,deSouzaDutra:2008gm, Peyravi:2021bra}, while others focus on coupled field potentials which lead to an Ising or Bloch-type domain wall inspired by ferromagnetic systems and condensed matter physics \cite{81,86,87,88,Peyravi:2015bra,Peyravi:2015kja}. Recently, solitons in the form of deformed kinks with power-law and exponential tails have become an interesting subject of research with important applications in physics. Currently, the necessary conditions for the existence and stability of kink solutions with power-law tails have been well studied \cite{kkka,lrik,kaka,DeG,a,b,d,e,f}. These new solutions present new characteristics and properties as compared with standard solitons. They have different stability conditions and more complicated dynamics, due to the long-range interaction in kink-antikink and kink-kink configurations. Indeed, applying appropriate deformations in the model provide substantially critical changes in the characteristics of kink solutions. In this way we are able to control important features of localized solutions, such as the kink amplitude, its rest mass and spatial scale, without any changes in the overall topological conditions. Using this method, one can also construct new topologies for the geometry of solitons \cite{g}, or transform non-topological lump-like structures of a specific field theory into a model supporting topological solutions \cite{h,i}.

The most important differences between these two types of kinks is that the energy density of the kink with an exponential tail is more localized in comparison with the power-law tail one. Thus, one may expect that for the power-law tail kink a large part of the soliton mass is concentrated in the tail \cite{kaka}. For instance, it has been proven that for any kink with a power-law tail, the stability potential decreases as a function proportional to the square inverse of the spatial position \cite{DeG}. These new generation of kinks have been produced as deformed solutions via a strictly monotonic deforming function with finite and infinite derivatives. In both cases, the exponential asymptotic of the kink remains exponential, and the power-law asymptotic remains power-law, although the numerical coefficient related to the length coordinate/space parameter may or may not vary. These changes, in turn, lead to properties that will affect the interaction of solitons and the force between them. Furthermore, it has been shown that the mass of kink changes, by taking into account the properties of the deforming function \cite{DeG}. 
The asymptotic behavior of the kink stability will directly affect the brane dynamics and its properties. The interaction between branes also can be studied through kink scattering, in particular, it has been shown that the interaction force between kinks with exponential and power-law tails (and thus between related branes) are weaker in comparison with symmetric kinks and it decreases faster as the distance between the kinks increases. 

The linear and nonlinear symmetries of the model depend on the characteristics of the potential $V(\phi)$ and its localized solutions. Indeed, the properties of a full effective action containing solitonic 5-branes are highly dependent on the type of the potential considered \cite{AA}. In particular, one of the most interesting representations of the early stages of the Universe is provided by D-branes, where the properties of the embedded D-brane are directly reflected in the representation of the Universe phase transitions and its potential energy. On the other hand, D-brane dynamics are related to the geometric representation of soliton objects in the nonlinear field added to the action \cite{ ALB}. It has been shown that, asymmetric thick branes can be obtained from both symmetric and asymmetric field potentials \cite{BM}. Indeed, new perspectives of asymmetric features can be added to the brane properties through field potentials with localized asymmetric solutions. Here, we consider potentials with asymmetric solutions in the forms of exponential and power-law tails. 

Although the $\phi^{4} $ model is a reputed theory of phase transitions and is based on the 
foundation and development of nonlinear physics, higher order-field theories with the possibility of 
changing the number of equilibria leads to more types of mesons on the one hand and long-range 
interactions between massless mesons on the other hand \cite{gir0,gir1}.
Furthermore,  successive phase transitions which drive the late time expansion of the universe, as well as the explicit connection between particle physics and the possibility of reconstructing the potential in 
colliders can be interpreted via these theories \cite{gir3, HiOr}. In addition to this, $\phi^{2n} $ field 
theories have an application as toy models for dark matter halos by considering them as Lane-
Emden truncations \cite{gir2}.

In this work, we are interested in exploring the effects of different exponential tail kinks as well as kink solutions with power-law tails of higher order field theory models ($\phi^{10}$ and $\phi^{18}$) in the braneworld scenario. The potential of exponential tail kinks has been constructed so that five/three distinct degenerate vacua in the $\phi$ field exist; on the other hand the potential of power-law tails generate only three distinct degenerate vacua in the $\phi$ field. Therefore, brane solutions appear due to the vacuum structure of the $\phi$ field and properties of tails.

The paper is outlined in the following manner: In Sec. \ref{11}, we present the general formalism of the braneworld scenario, by writing the action and gravitational field equations, and analyse the particle motion near the brane through the geodesic equations. In  Sec. \ref{22}, we introduce higher order field theory models ($\phi^{10}$ and $\phi^{18}$) and present their soliton solutions. Furthermore, we analyse in detail the behaviour of the superpotentials, the warp factors, the Ricci and Kretschmann scalars and the mixed Einstein tensor components. In Sec. \ref{33}, we explore the stability regions of the potential of the linearised Schr\"{o}dinger equation as a function of the free parameters of the model. Finally, in Sec. \ref{44}, we conclude.

\section{Thick Brane Formalism}\label{11}

We consider a thick brane, embedded in a five-dimensional (5D) bulk spacetime, modelled by the following action:
\begin{equation}
S=\int
d^{5}x\sqrt{|g^{(5)}|}\left[\frac{1}{4} R[g^{(5)}]-\frac{1}{2}\partial_{A}\phi\partial^{A}\phi-V(\phi)\right],
\end{equation}
where we use the notation $k_{5}^{2}=8 \pi G_{5}=2$; $R[g^{(5)}]$ is the scalar curvature in the bulk, and $g^{(5)}$ is the metric, $\phi$ is a dilaton bulk field and $V(\phi)$ is a general potential energy.

Consider the following line element of the brane, embedded in the 5D bulk spacetime \cite{br}:
\begin{eqnarray}
ds^{2}_{5}&=&g_{CD}dx^{C}dx^{D}\nonumber\\
&=&e^{2A}\eta_{\mu\nu}dx^{\mu}dx^{\nu}+dw^{2},
\end{eqnarray}
with the metric signature $(-,+,+,+,+)$, and where $C, D=0...4$, $\mu, \nu=0...3$; $A$ is the warp factor which depends only on the 5D coordinate $w$. The 5D energy-momentum
tensor $T_{CD}=\partial_{C}\phi\partial_{D}\phi-g_{CD}\left[\frac{1}{2}\partial_{B}\phi\partial^{B}\phi+V(\phi)\right]$
is defined in terms of the scalar field $\phi$ with the potential $V(\phi)$, where $g_{CD}$ and $\phi$ are functions solely of $w$. Thus, the 5D equations of motion are given by:
\begin{eqnarray}
\label{e1}
3A^{\prime\prime}+6{A^{\prime}}^{2}&=&-k_{5}^{2}e^{-2A}T_{00}=-k_{5}^{2}\left[\frac{1}{2}{\phi^{\prime}}^{2}+V(\phi)\right] ,
     \\
\label{e2}     
6{A^{\prime}}^{2}&=&k_{5}^{2}T_{44}=k_{5}^{2}\left[\frac{1}{2}{\phi^{\prime}}^{2}-V(\phi)\right] , 
     \\
\label{e3}      
\phi^{\prime\prime}+4A^{\prime}\phi^{\prime}&=&\frac{d
V(\phi)}{d\phi},
\end{eqnarray}
where the prime denotes a derivatives with respect to $w$.

However, it is more practical to reduce these Einstein-scalar field equations into first-order equations, by introducing an auxiliary function $W$ \cite{de,Sa,Af,CM,DA}. In this regard the scalar field potential $V(\phi)$ is given by:
\begin{equation}\label{vti}
V(\phi)=\frac{1}{8}\left(\frac{\partial
W(\phi)}{\partial\phi}\right)^{2}-\frac{1}{3}W(\phi)^{2}.
\end{equation}
Then, it is easy to show that the above equations become:
\begin{eqnarray}
\label{ap2}
\phi^{\prime}&=&\frac{1}{2}\frac{\partial
W(\phi)}{\partial\phi},\\
\label{ap}
A^{\prime}&=&-\frac{1}{3}W(\phi), 
\end{eqnarray}
The scalar field as a function of the fifth dimension ($\phi(w)$), the auxiliary function $W$ and the warp factor $A$ are obtained by these equations respectively.
Using these definitions, the energy density distribution on the bulk, $T_{00}$, which will be analysed in detail below, is given by \cite{blg}:
\begin{equation}
T_{00}=e^{2A}\left[\frac{1}{2}\left(\frac{\partial\phi}{\partial
w}\right)^{2}+V(\phi)\right].
\end{equation}

For models with an infinitely thin brane and Dirac delta distributions, the energy density is equal to the cosmological constant of the bulk plus the energy density on the brane, i.e., $\varepsilon=\Lambda_{5}^{\pm}+k\delta(w)$, where $k$ is parameter independent of $w$ and is related to the energy density on the brane \cite{Peyravi:2015bra,Peyravi:2015kja}, where $\Lambda$ is a cosmological constant on the brane, which could be positive, negative or zero corresponding to the $4D$ spacetime being de Sitter ($dS_{4}$), anti-de Sitter ($AdS_{4}$) or Minkowski ($M_{4}$), respectively \cite{Sa,Af}.

Moreover, it is also interesting to calculate the geodesic equation along the fifth dimension in a thick brane, in order to investigate the particle motion near the brane \cite{JS}. Thus, the geodesic equation yields the following relation (we refer the reader to \cite{Peyravi:2015bra,Peyravi:2015kja} for more details): $\ddot{w}-c_{1}^{2}\left[f(w)\right]=0$, where $f(w)$ is defined as $f(w)=A'(w)e^{-2A(w)}$ and $c_{1}$ is a constant of integration. Note that the solution of $w$ depends critically on whether $f(w)$ is positive or negative. More specifically, for positive (negative) values of $f(w)$ one obtains exponential (periodic) solutions, respectively. The exponential solutions indicate that the reference point is unstable, while the periodic motion indicates particle confinement near the brane. 
In the periodic situation, by introducing the following new function: $F(w)=-c_{1}^{2}A'(w)e^{-2A(w)}$, the geodesic equation assumes the simplified form $\ddot{w}+F(w)=0$. One can show that in the exact stable point, i.e., $w_{0}$, that $F(w_{0})=0$. Otherwise, expanding $F(w)$ around $w_{0}$, we have $F(w)=F(w_{0})+F'(w_{0})(w-w_{0})+...$, and the geodesic equation takes the form $\ddot{w}+F'(w_{0})(w-w_{0})=0$. For simplicity, using a change of variable given by $\tilde{w}=w-w_{0}$, the geodesic equation reduces to $\ddot{\tilde{w}}+\Omega^{2}(\tilde{w})=0$, where $\Omega=\sqrt{F'(w_{0})}$ \cite{Peyravi:2015bra}.
It is interesting to note that in RS-2 brane model, the $KK$ zero mode corresponds to the massless graviton and the massive modes form a continuum which result in a small correction to Newtonian gravity at large distances \cite{YZYX,FDCR}. Note that free particles are affected by the gravitational field and not directly by the scalar field, while any field or particle directly coupled with the scalar field, will be further affected by an extra force arising from the scalar field. In particular, an exponential potential, such as $\exp{(-\alpha w)}$, reduces to a harmonic potential for small amplitude oscillations ($e^{-\alpha w}\approx1-\alpha w^{2}+O(w^{4})$).      

In the following section, we explore several models for thick branes and their characteristics.

\section{Thick Brane with Exponential and Power-Law Tails}\label{22}

In this section, we review three higher-order self-interaction potentials and their soliton solutions in flat space-time \cite{HiOr}, in order to study the brane world scenario of these models by considering the above-mentioned kink solutions as thick branes in curved space-time. It should be noted that these three models include two explicit kink solutions with exponential tails via $5$ and $3$ degenerate minima potential (ET 5) and (ET 3), respectively, and one with power-law tail (PT).

Thus, we start our investigation by considering the following self-interaction potentials in the flat space-time \cite{HiOr}:
\begin{eqnarray}
\label{p1}
\tilde{V}_{ET 5}(\phi)&=&\lambda ^2 \phi ^2 \left(a^{2 n}-\phi ^{2 n}\right)^2 \left(b^{2 n}-\phi ^{2 n}\right)^2,\\  
\label{p2}
\tilde{V}_{ET 3}(\phi)&=&\lambda ^2 \phi ^2 \left(a^{2 n}-\phi ^{2 n}\right)^2 \left(b^{2 n}+\phi ^{2 n}\right)^2,\\ 
\label{p3}
\tilde{V}_{PT}(\phi)&=&\lambda ^2 \phi ^{2 (n+1)} \left|\left(a^{2 n}-\phi ^{2 n}\right)\right|^{3},
\end{eqnarray}
where $a$, $b$ and $\lambda $ are constants; $n$ is an integer number which starts from 1. The parameters $a$ and $b$ show the location of vacua in the potential, while the parameter $\lambda$ controls the depth and height of the potential extremes. On the other hand, $\lambda$ corresponds to the brane thickness, such that for $\lambda\rightarrow\infty$, the domain wall reduces to the step function and the energy density approaches a $\delta$-function.

\begin{figure*}[th]
\epsfxsize=9cm\centerline{\hspace{10cm}\epsfbox{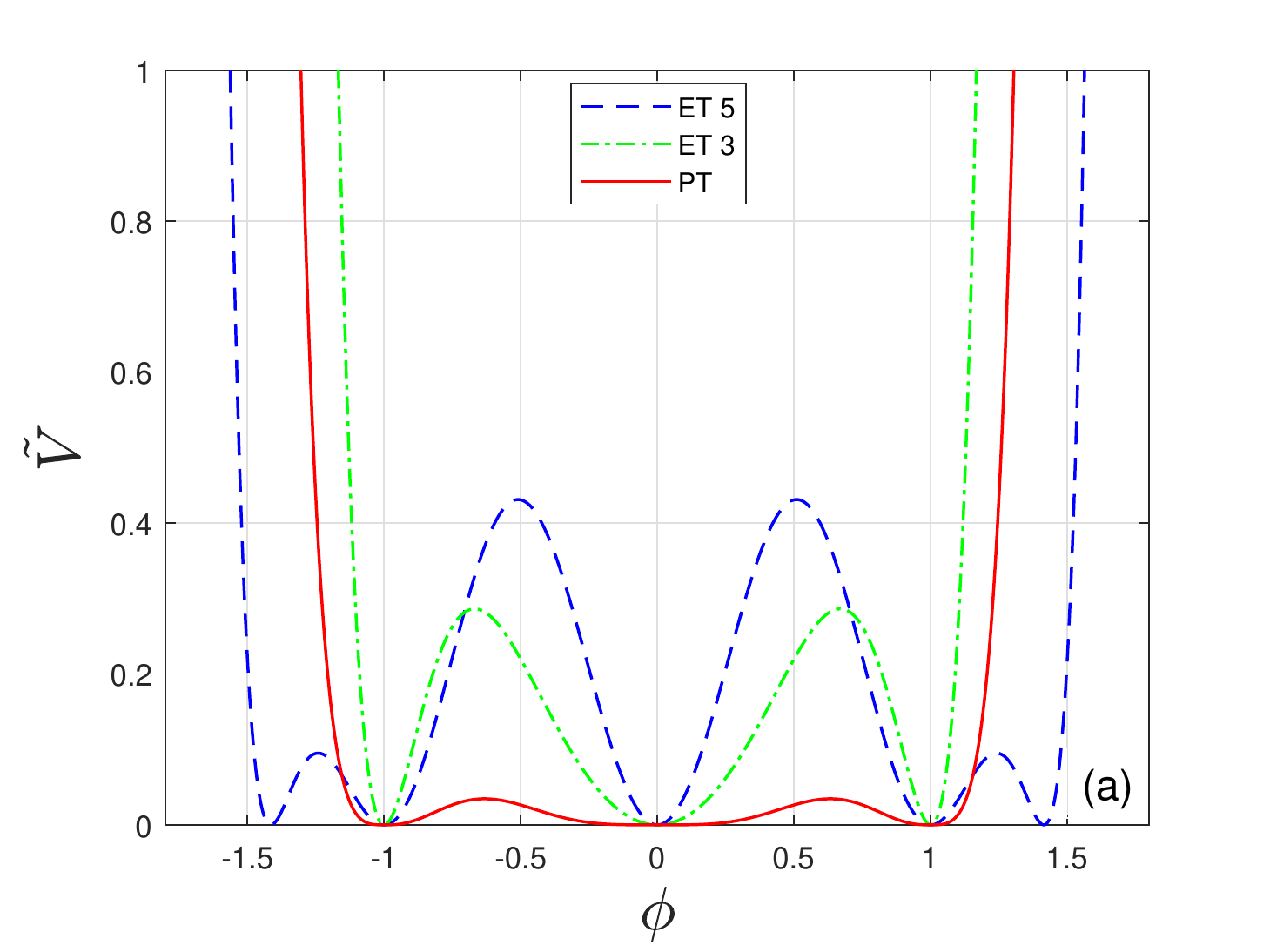}\epsfxsize=9cm\centerline{\hspace{-10cm}\epsfbox{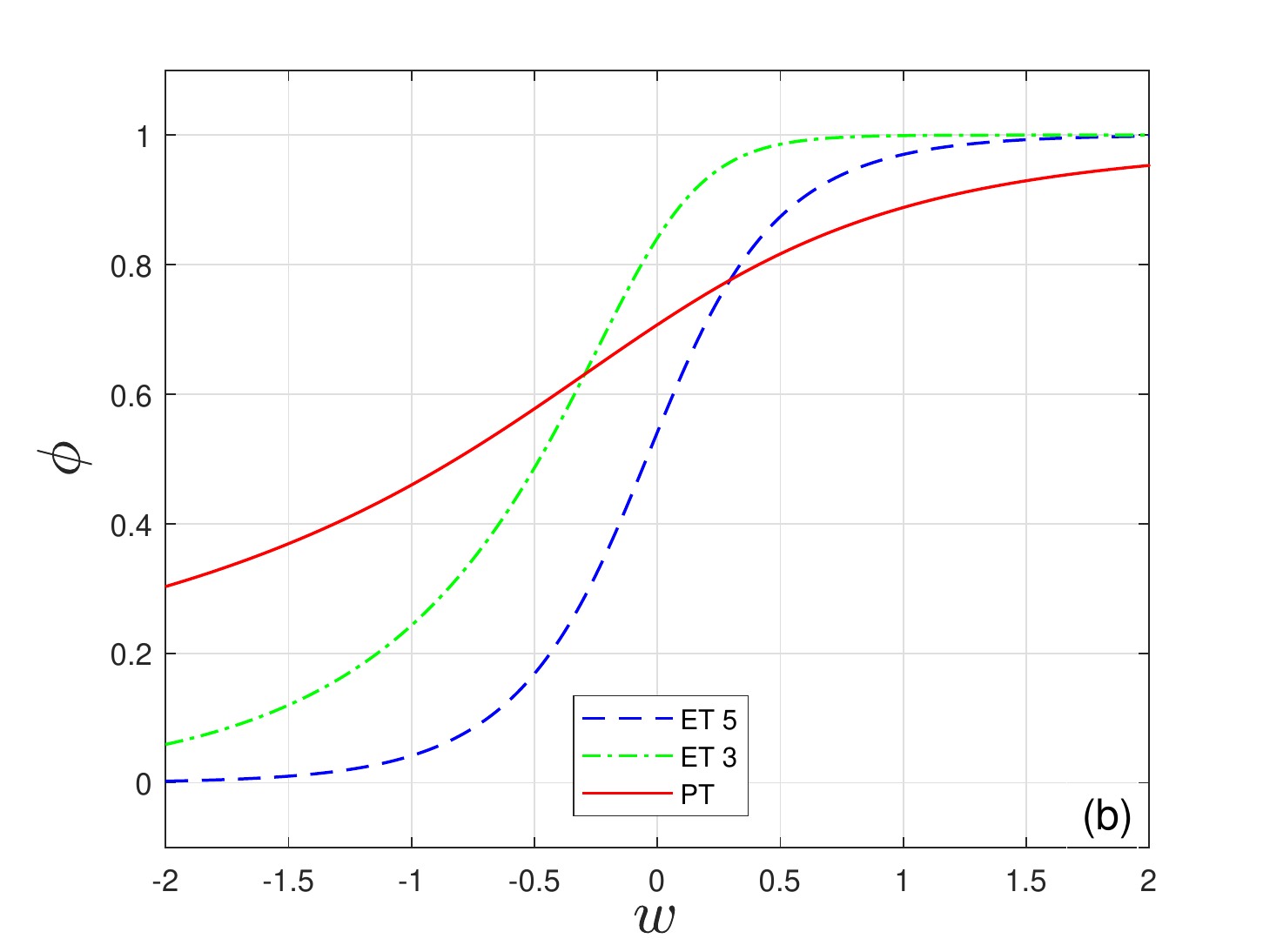}}}
\caption{The plots depict: (a) the potential $\tilde{V}$ as a function of $\phi$ and (b) the soliton solution $\phi$ as a function of $w$ for $n=1$ ($\phi^{10}$) and $\lambda=a=1$.  The dashed, dotted-dashed, and continuous curves correspond to ET 5 (Eq. (\ref{p1}), Eq. (\ref{f1})), ET 3 (Eq. (\ref{p2}), Eq. (\ref{f2})) and PT (Eq. (\ref{p3}), Eq. (\ref{f3})) respectively.  \label{Vtphin1}}
\end{figure*}

\begin{figure*}[th]
\epsfxsize=9cm\centerline{\hspace{10cm}\epsfbox{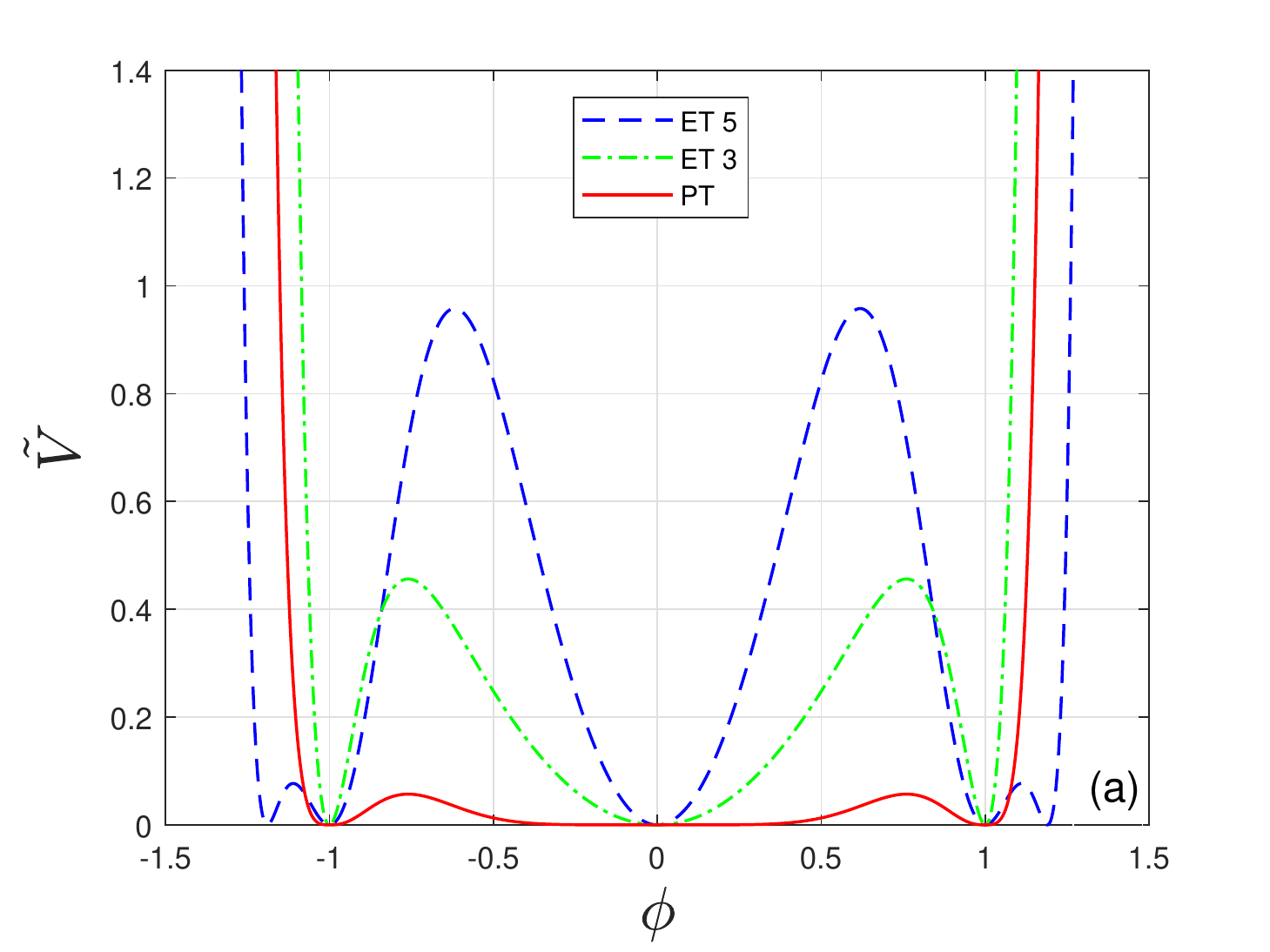}\epsfxsize=9cm\centerline{\hspace{-10cm}\epsfbox{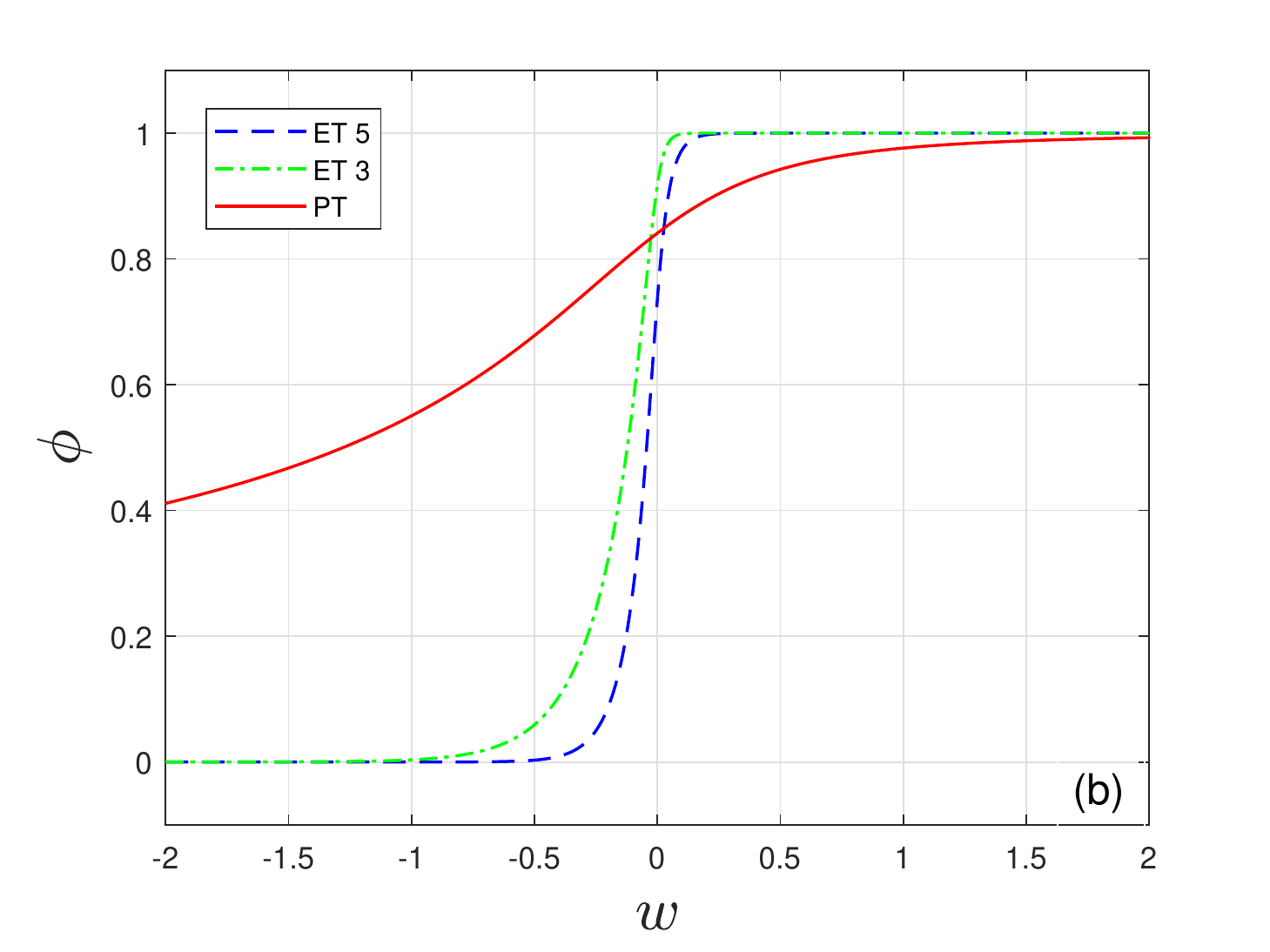}}}
\caption{The plots depict the following:
(a) the potential $\tilde{V}$ as a function of $\phi$ and (b) the soliton solution $\phi$ as a function of $w$ for $n=2$ ($\phi^{18}$) and $\lambda=a=1$.  Dashed, dotted-dashed, and continuous curves correspond to ET 5 (Eq. (\ref{p1}), Eq.  (\ref{f1})), ET 3 (Eq. (\ref{p2}), Eq. (\ref{f2})) and PT (Eq. (\ref{p3}), Eq. (\ref{f3})) respectively.\label{Vtphin2}}
\end{figure*}

One can easily verify that the potential (\ref{p1}) possesses $5$ degenerate minima at $\phi=0, \pm a, \pm b$ (with $b>a>0$), that lead to two kink and two mirror kinks (with the related corresponding four anti-kinks). On the hand, the potentials (\ref{p2}) and (\ref{p3}) have $3$ degenerate minima which are located in $\phi=0, \pm a$, which lead to one kink, one mirror kink and two anti-kinks \cite{HiOr}. We note that $n=1,2,3,...  $ correspond to order $\phi=10,18,26,...$ respectively, however, among these $n$, we will only focus on $n=1,2$ (order $\phi=10,18$). Figures \ref{Vtphin1}.a) and \ref{Vtphin2}.a) shows these potential for $n=1,2$. 

As mentioned before, each potential has different topological sectors, where each sector contains several explicit kink solutions that correspond to the specific constant in the potentials. However, for the cases $b^{2 n}=2 a^{2 n}$ and $b=a$ (for Eqs. (\ref{p1}) and (\ref{p2}), respectively) and in the sector $0$ to $a$, the following explicit solutions are obtained \cite{HiOr}:
\begin{align}
\label{f1}
\phi_{ET 5}(w)& = a\left[1-\frac{1}{\sqrt{e^{S_{n} w}+1}}\right]^{1/2n},\\ 
\label{f2}
\phi_{ET 3}(w)&= a\left[\frac{e^{S_{n} w}}{1 + e^{S_{n} w}}\right]^{1/4n},\\  
\label{f3}
\phi_{PT}(w) &=\frac{a}{2^{1/2 n}}\left[1+\frac{S_{n}  w}{\sqrt{S_{n} ^2 w^2+64}}\right]^{1/2n}, 
\end{align}
where $S_{n}$ is defined as:
\begin{equation}
S_{n}=4\sqrt{2}n\lambda a^{4n}.
\end{equation}
These localized kinks with exponential, Eqs. (\ref{p1}) and (\ref{p2}), and power-law tails, Eq. (\ref{p3}), are represented in Figs. \ref{Vtphin1}b and \ref{Vtphin2}b.

Since we want to represent the kink solutions as thick branes in curved space-time via Eqs. (\ref{e1})-(\ref{e3}), the potential ($\tilde{V}$) should be modified through Eq. (\ref{vti}).
Besides, there is a spatial relationship between the derivative of the scalar field with respect to the fifth dimension ($\phi^{\prime}$) and the derivative of the auxiliary function ($W$) with respect to the scalar field ($\phi$) (Eq. (\ref{ap2})). While, according to the Eq. (\ref{ap}) the derivative of the warp factor with respect to the fifth dimension ($A^{\prime}$)  is related to the auxiliary function ($W$). Therefore, it is possible to find the specific $W$ regarding to the given $\tilde{V}$ and $\phi$ in flat space-time and consequently obtain the corresponding potential $V$ in the curved space-time via Eq. (\ref{vti}).


Thus, in order to study these brane scenarios, the first step is to obtain the warp factor, $A$. However, the mathematical form of the self-interaction potential $\tilde{V}$ and the kink solution $\phi$ of the considered systems are too complicated to find $A(w)$, by above-mentioned process. Thus, solving Eqs. (\ref{ap2}) and (\ref{ap}) analytically for higher order systems is a highly non-trivial issue. 
In this regard, one can reduce the real scalar field equation ($\Box \phi = - d \tilde{V}/d \phi$), for the static case to $d^2\phi/dw^2= d \tilde{V}/d \phi$  \cite{gui,DeG} which, in turn, reduces to the following first order ordinary differential equation:
\begin{equation}\label{psam}
\frac{d\phi}{dw}=\phi^{\prime}=\pm\sqrt{2\tilde{V}}.
\end{equation}
The next step is to combine Eq. (\ref{ap2}) and Eq. (\ref{psam}), which yields:
\begin{equation}\label{sam}
W=2\sqrt{2}\int{\sqrt{\tilde{V}}d\phi},
\end{equation}
which is a straightforward manner for obtaining the auxiliary function $W$ from the self-interaction potential $\tilde{V}$. Thus, taking into account Eqs. (\ref{ap2}) and (\ref{ap}), one can write $A$ as:
\begin{equation}\label{sam2}
A=-\frac{1}{3\sqrt{2}}\int \frac{W}{\sqrt{\tilde{V}}}d\phi.
\end{equation}
Thus, by solving Eq. (\ref{sam}) for the auxiliary function $W(\phi)$ for the three models considered, the warp factor can be calculated from Eq. (\ref{sam2}).

\begin{figure*}
\epsfxsize=9cm\centerline{\hspace{10cm}\epsfbox{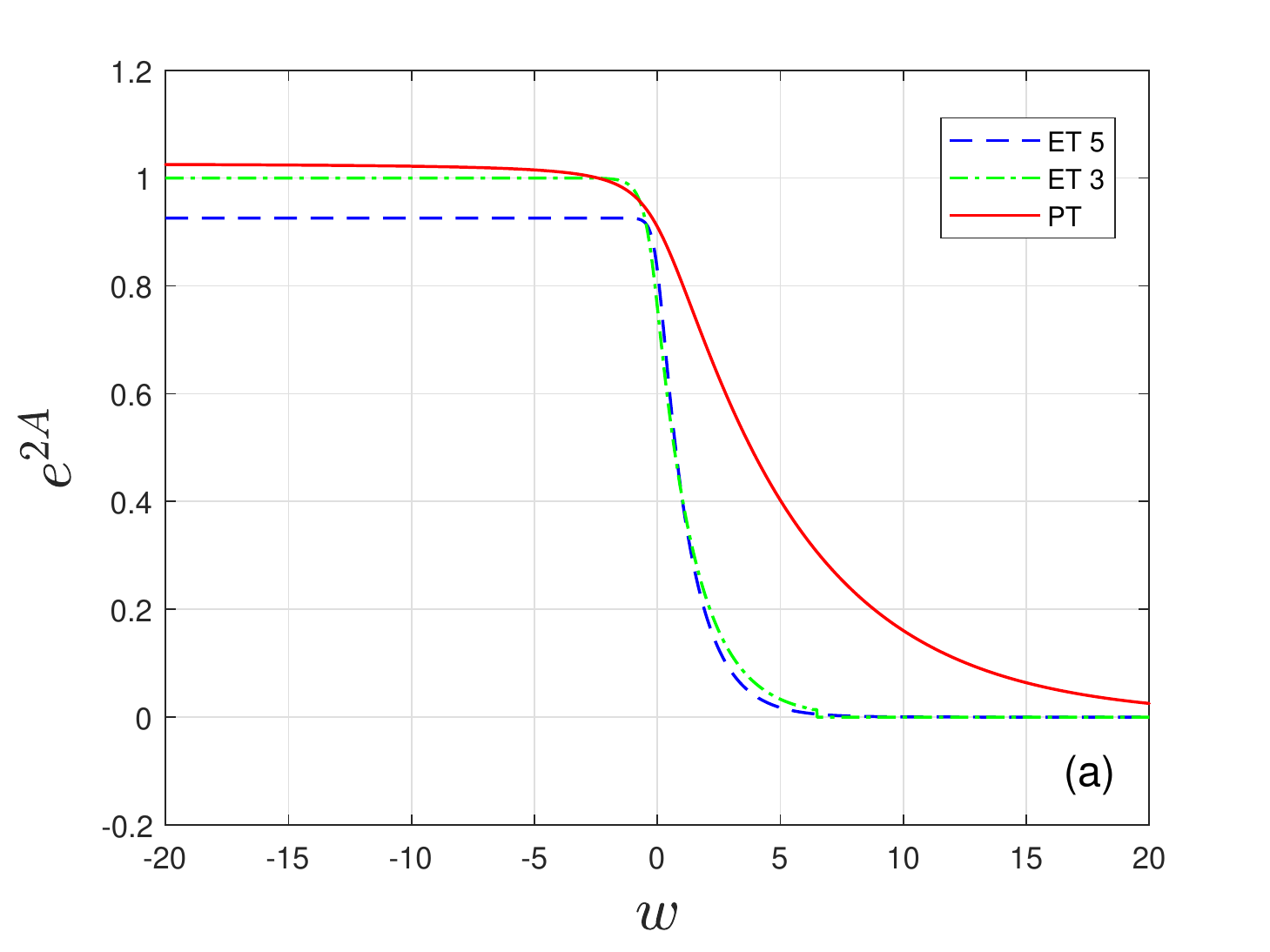}\epsfxsize=9cm\centerline{\hspace{-10cm}\epsfbox{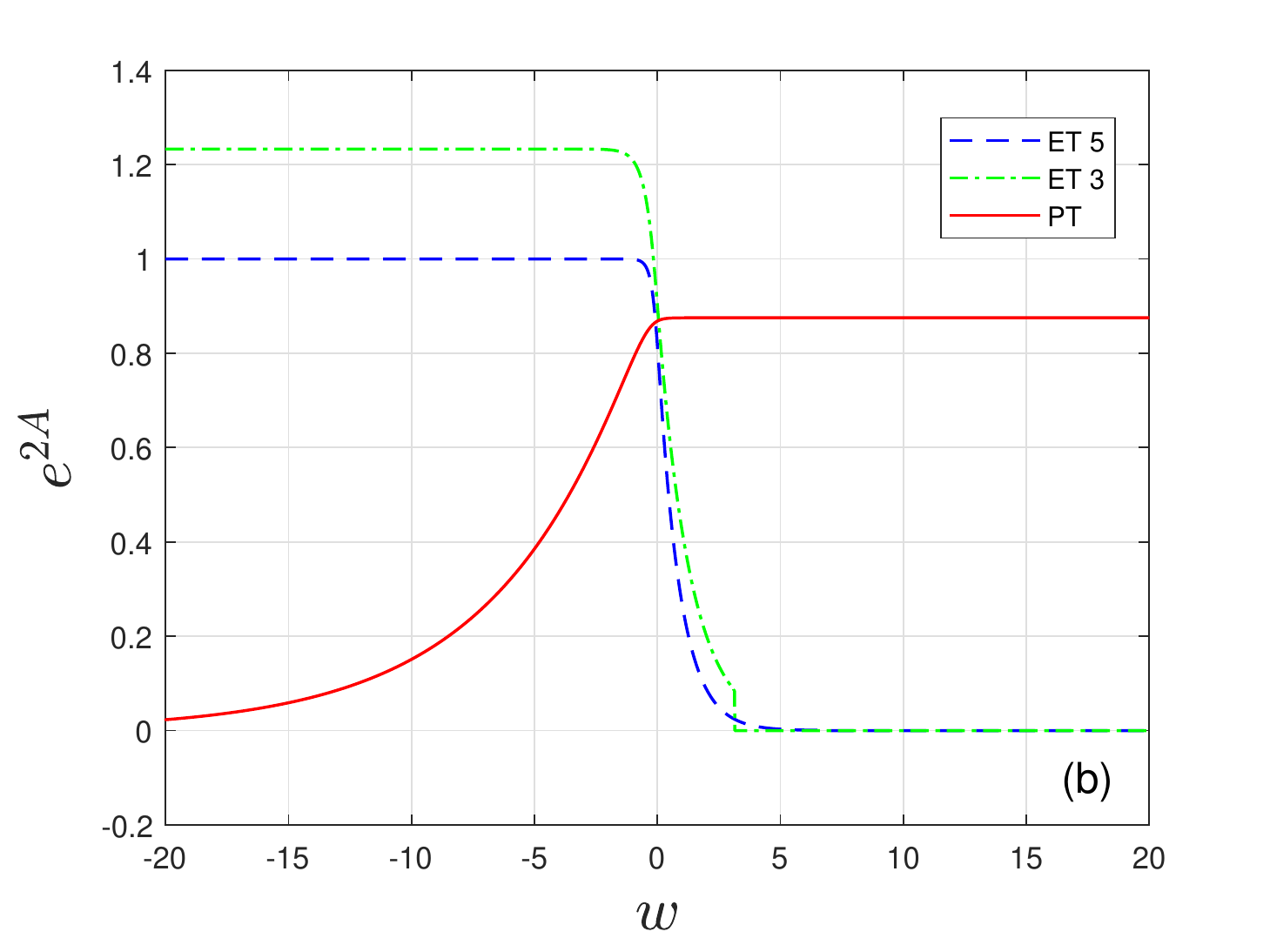}}}
\caption{The plots depict the warp factor as a function of the fifth dimension for the ET 5, ET 3 and PT cases for $\lambda=a=1$.
(a) $n=1$  ($\phi^{10}$). Dashed, dotted-dashed, and continuous curves correspond to ET 5 (Eq. (\ref{Ap1})), ET 3 (Eq. (\ref{Ap9})) and PT (Eq. (\ref{Ap19})) respectively; (b) For $n=2$ ($\phi^{18}$), the dashed, dotted-dashed, and continuous curves correspond to ET 5 (Eq. (\ref{Ap5})), ET 3 (Eq. (\ref{Ap13})) and PT (Eq. (\ref{Ap23})), respectively.\label{wp}}
\end{figure*}
\begin{figure*}[h]
\epsfxsize=9cm\centerline{\hspace{10cm}\epsfbox{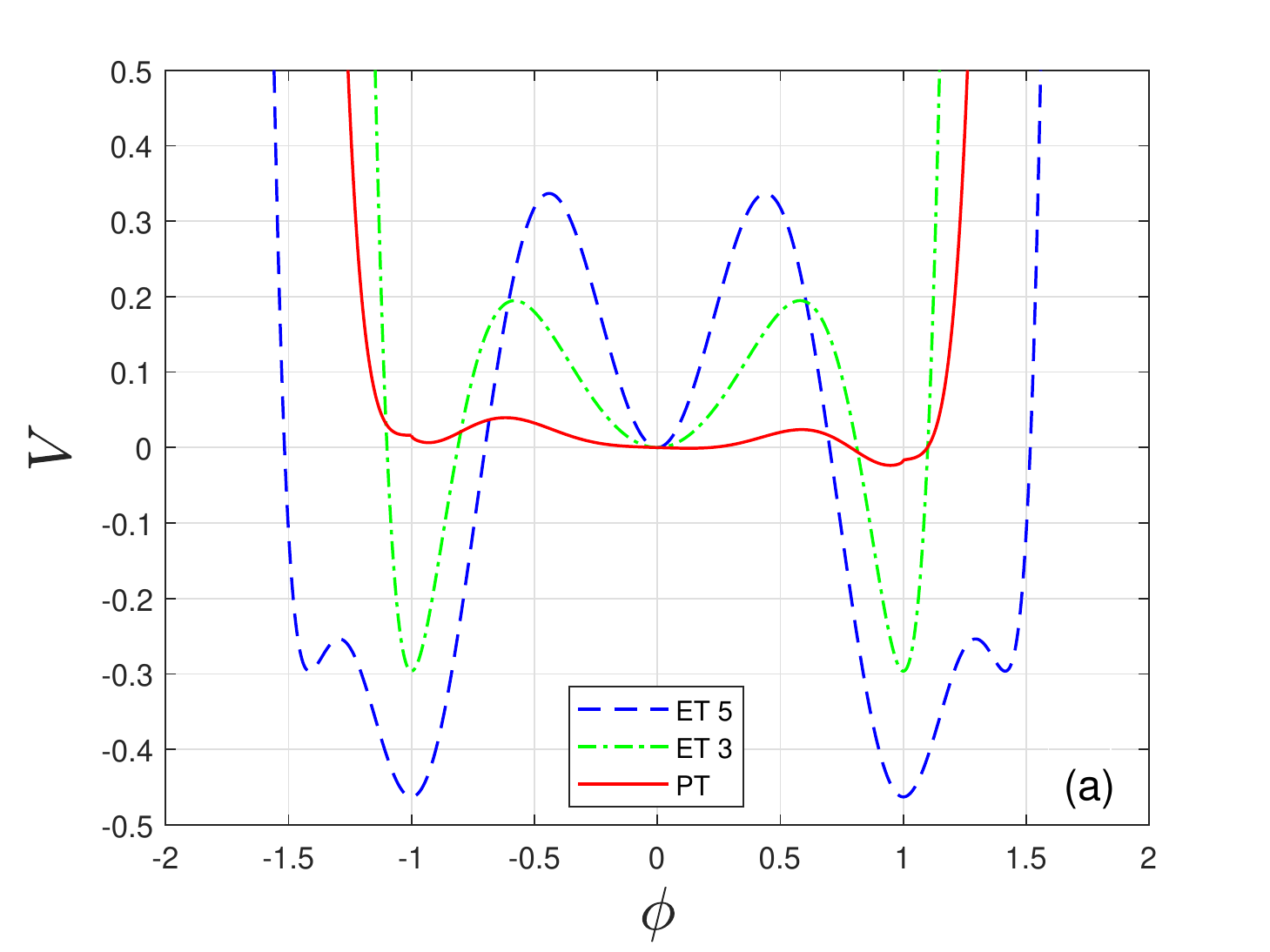}\epsfxsize=9cm\centerline{\hspace{-10cm}\epsfbox{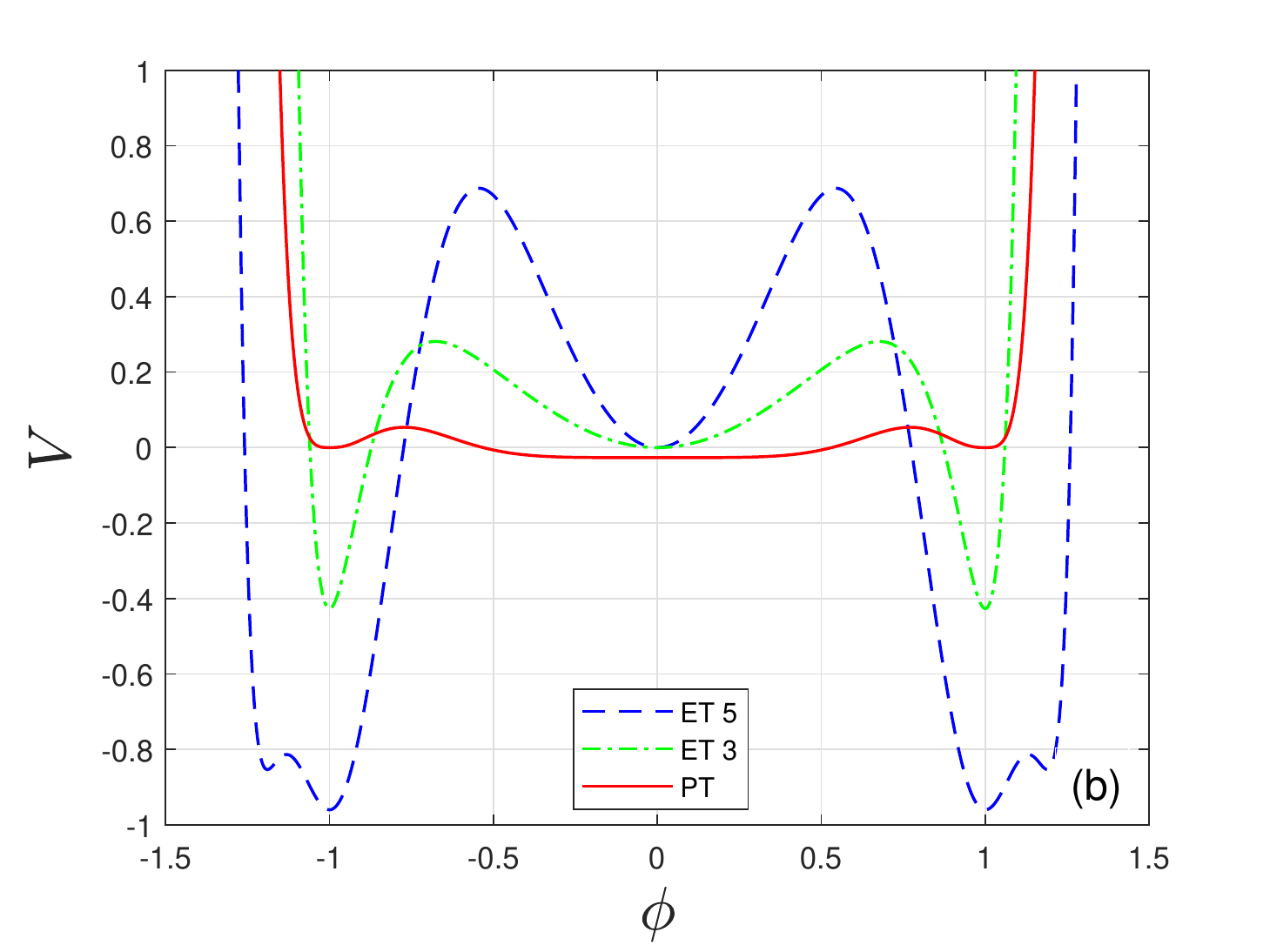}}}
\caption{The plots depict the potential $V$ as a function of $\phi$ for Eq. (\ref{V5}), Eq. (\ref{V3}) and  Eq. (\ref{Ap18}) for $\lambda=a=1$. (a) for $n=1$ ($\phi^{10}$) and (b) for $n=2$ ($\phi^{18}$), Eq. (\ref{V5}), Eq. (\ref{V3}) and  Eq.(\ref{Ap23a}) respectively.\label{Vphin12}}
\end{figure*}

In order to investigate brane scenarios for the models considered in this work, all the relevant quantities have been calculated, namely the auxiliary function, $W$, the modified potential, $V$, the warp factor, $A$, Ricci scalar $R$, the mixed Einstein tensor components $G_{D}^{C}$, the energy momentum tensor $T_{00}$ and geodesic equation are presented in the Appendixes \ref{appendix:AA}-\ref{appendix:CC}. However, these equations are extremely lengthy and rather than writing down all the explicit analytical functions here, we summarize the results in the figures presented, and refer the interested reader to the Appendixes \ref{appendix:AA}-\ref{appendix:CC} regarding the three considered models, respectively, for the full solutions considered.

Figure \ref{wp} demonstrates the warp factors $A$ of orders $\phi^{10}$ and $\phi^{18}$, respectively. As one can readily verify, all of these curves increase from the left ($w=-\infty$) to right ($w=+\infty$), except the warp factor of $\phi^{18}$ brane via PT which manifests the inverse behaviour. The reason is that for all cases, the minimum of potential at $a$ point is lower than minimum at $0$, while this fact is the inverse for the $\phi^{18}$ PT case (See Fig. \ref{Vphin12}). Furthermore, as expected the warp factors of the exponential tail branes are more localized in comparison to the warp factors of the power-law tail branes.

Now, the next step in our analysis consists in obtaining the modified potential from Eq. (\ref{vti}). 
In contrast with the $\phi^{4}$ and $\phi^{6}$ cases, the modified potential of the models considered in this work (see Fig. \ref{Vphin12}) have the same number of minima with the original self-interaction potential \cite{Peyravi:2015bra}. However, for the ET 5 and ET 3 cases, the minima are non-degenerate which implies that these models lead to $Z_{2}$-symmetry breaking branes. On the other hand, the center of the brane position can play a vital role in $Z_{2}$-symmetry. In this regard, Fig. \ref{Vphin12} shows that for the ET 5 model the maximum of the modified potential is exactly located between vacuum $0$ and $a$. Furthermore, as mentioned above, the parameter $\lambda$ not only controls the depth and height of potential extremes, but it can also change the position of the maximum to some extent.

\begin{figure*}[th]
\epsfxsize=9cm\centerline{\hspace{10cm}\epsfbox{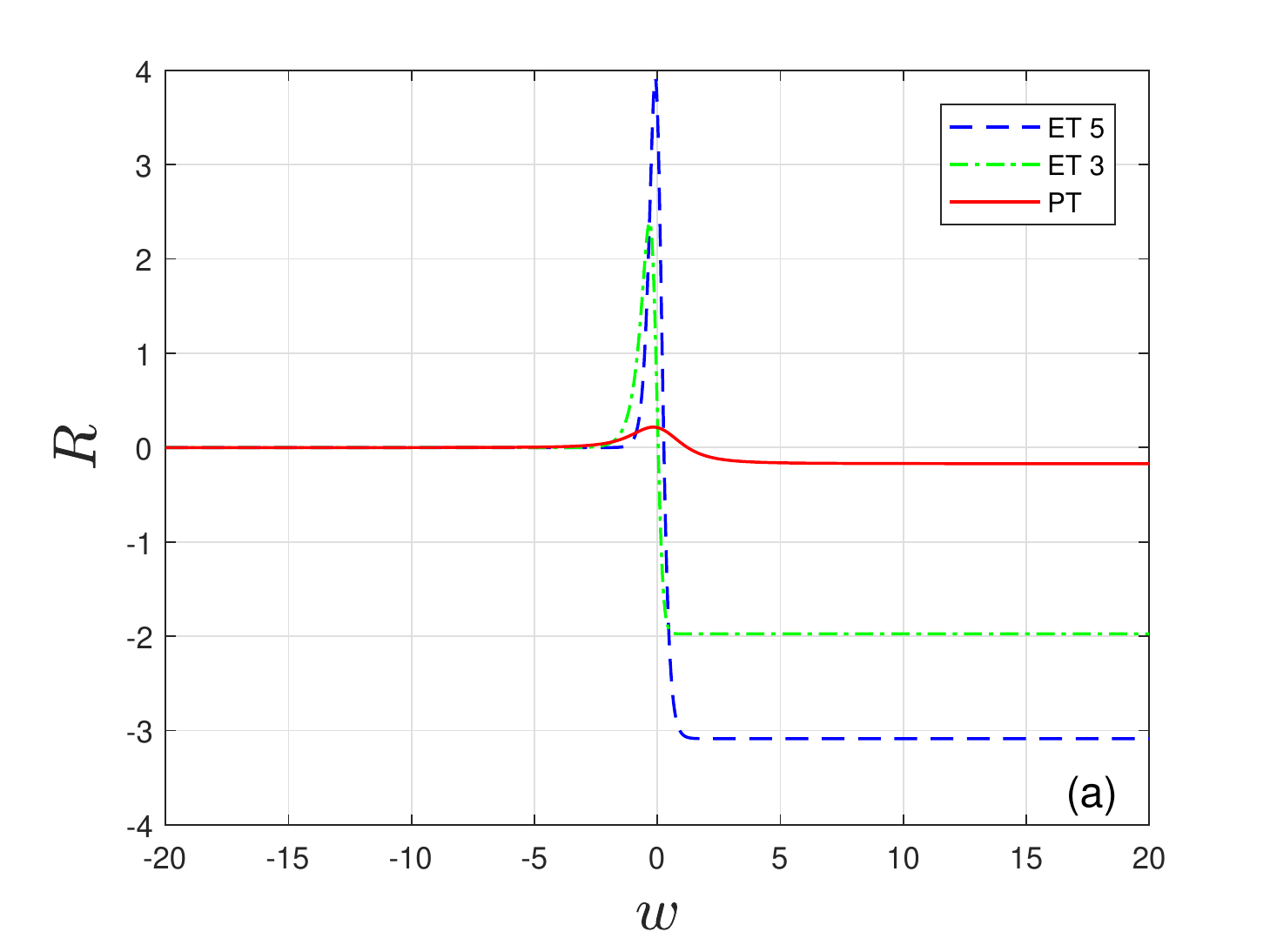}\epsfxsize=9cm\centerline{\hspace{-10cm}\epsfbox{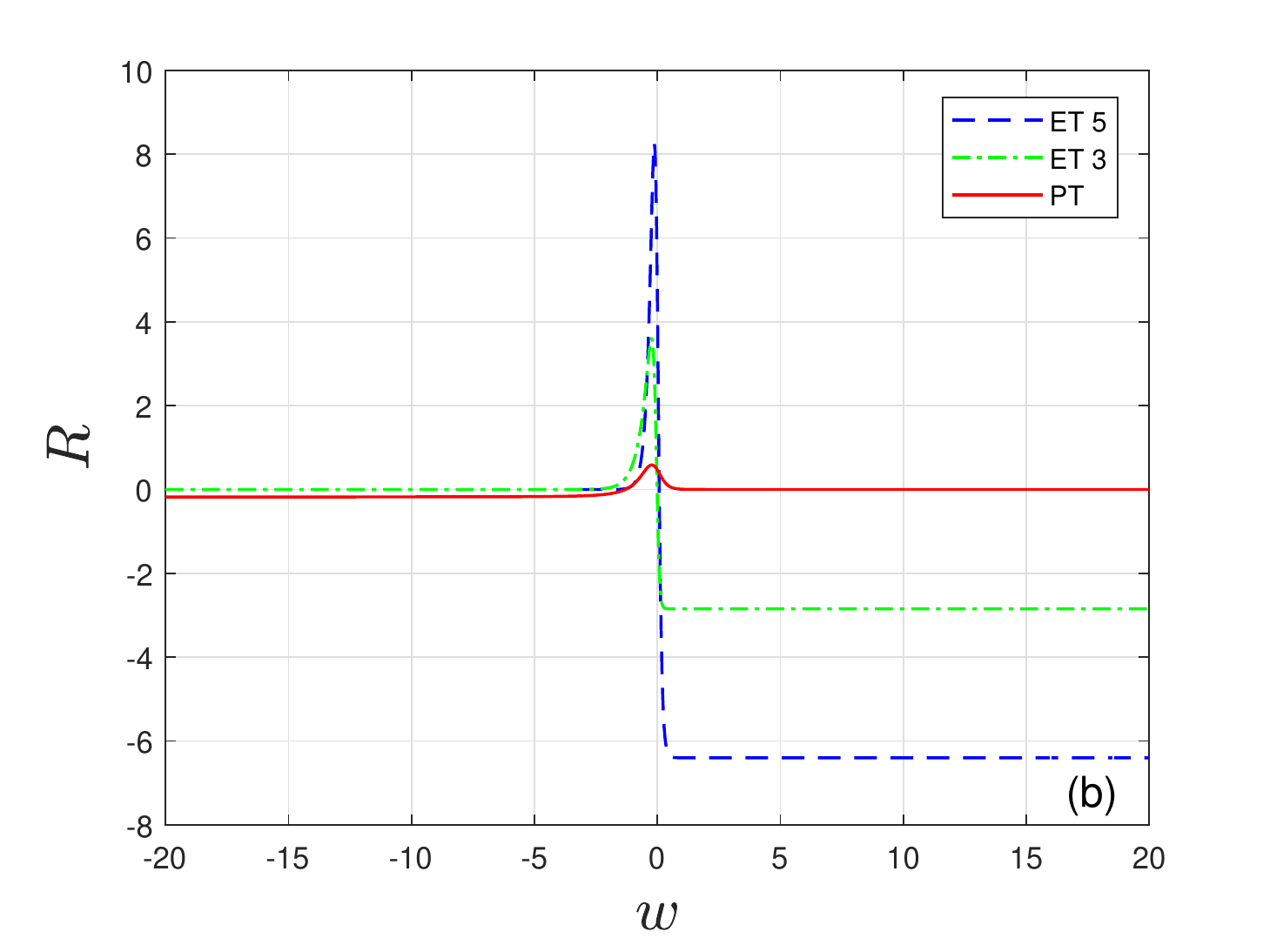}}}
\caption{The plots depict the Ricci scalar as a function of the fifth dimension coordinate $w$ for the ET 5, ET 3 and PT cases for $\lambda=a=1$.
(a) For $n=1$ ($\phi^{10}$),  the dashed, dotted-dashed and continuous curves correspond to ET 5 (Eq. (\ref{Ap2})), ET 3 (Eq. (\ref{Ap10}))  and PT (Eq. (\ref{Ap20})), respectively. (b) For $n=2$ ($\phi^{18}$),  the dashed, dotted-dashed and continuous curves correspond to ET 5 (Eq. (\ref{Ap6})), ET 3 ( Eq. (\ref{Ap14})) and PT (Eq. (\ref{Ap24})), respectively.\label{r}}
\end{figure*}

\begin{figure*}[th]
\epsfxsize=9cm\centerline{\hspace{10cm}\epsfbox{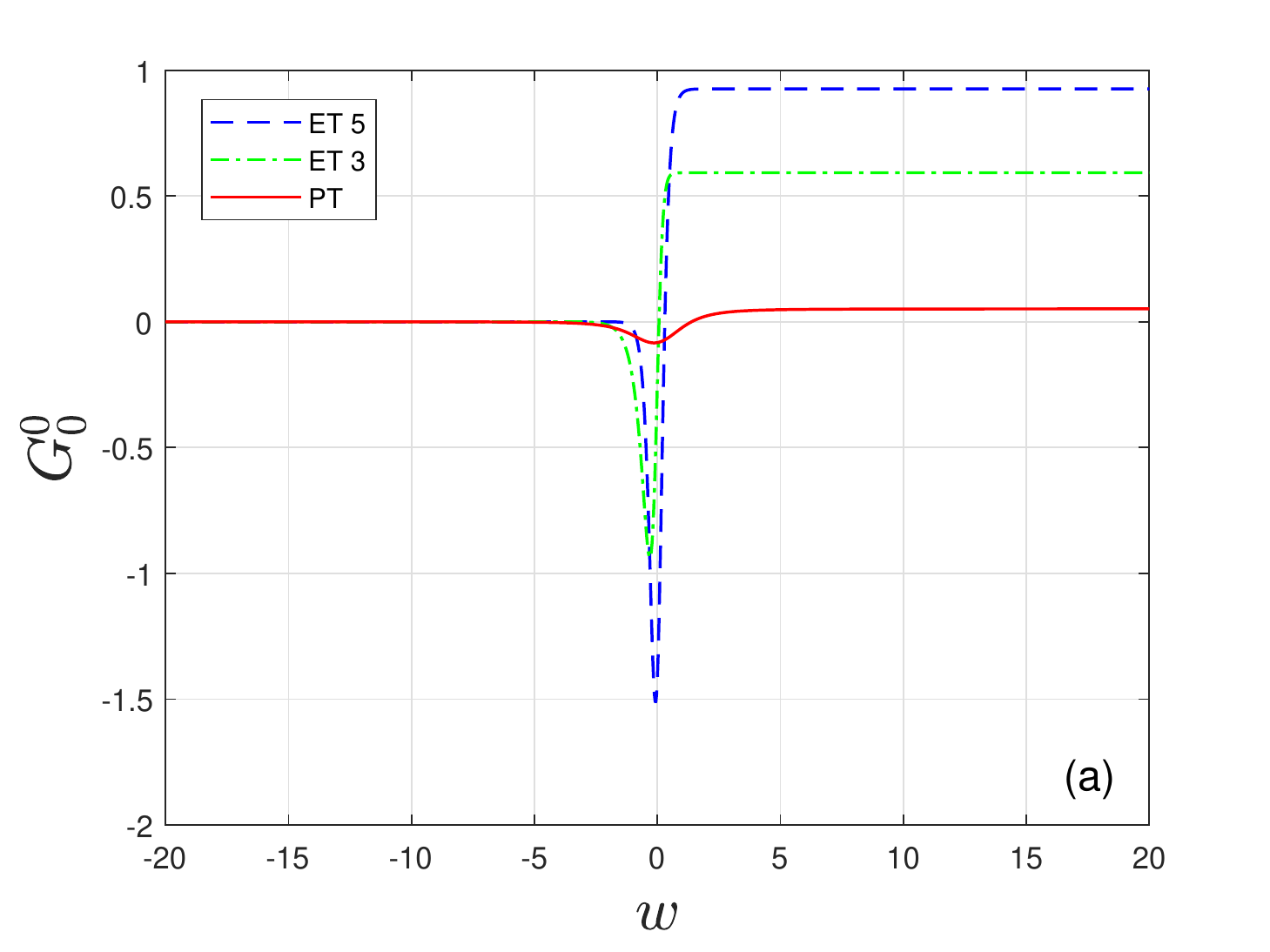}\epsfxsize=9cm\centerline{\hspace{-10cm}\epsfbox{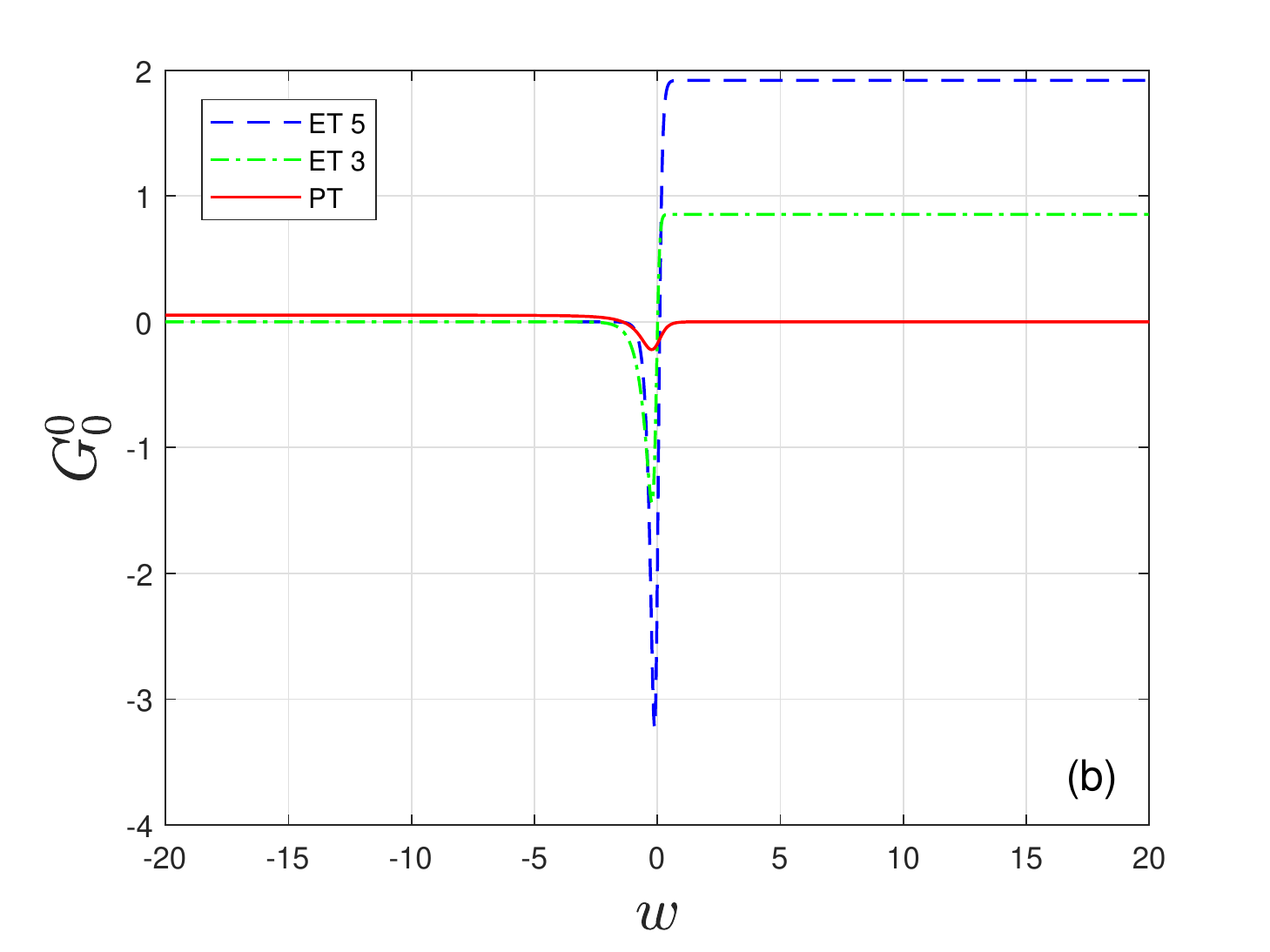}}}
\caption{The plots depict the Einstein tensor component $G_{0}^{0}$ a function of the fifth dimension coordinate $w$ for the the ET 5, ET 3 and PT cases for $\lambda=a=1$.
(a) For $n=1$ ($\phi^{10}$), the dashed, dotted-dashed and continuous curves correspond to ET 5 (Eq. (\ref{Ap3})), ET 3 (Eq. (\ref{Ap11})) and PT (Eq. (\ref{Ap21})), respectively. (b) For $n=2$ ($\phi^{18}$), the dashed, dotted-dashed and continuous curves correspond to ET 5 (Eq. (\ref{Ap7})), ET 3 (Eq. (\ref{Ap15})) and PT (Eq.(\ref{Ap25})), respectively.\label{gw}}
\end{figure*}

\begin{figure*}[th]
\epsfxsize=9cm\centerline{\hspace{10cm}\epsfbox{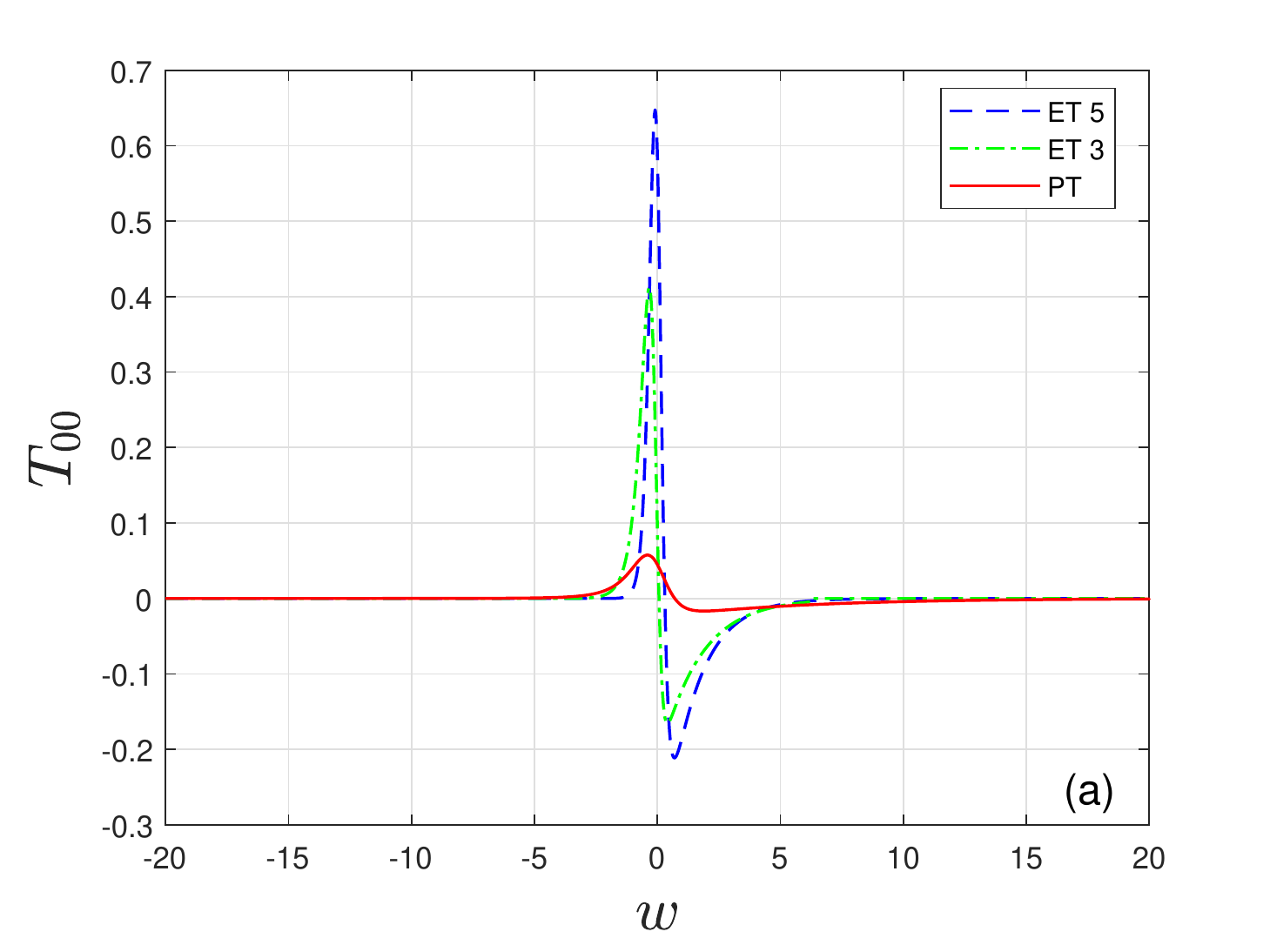}\epsfxsize=9cm\centerline{\hspace{-10cm}\epsfbox{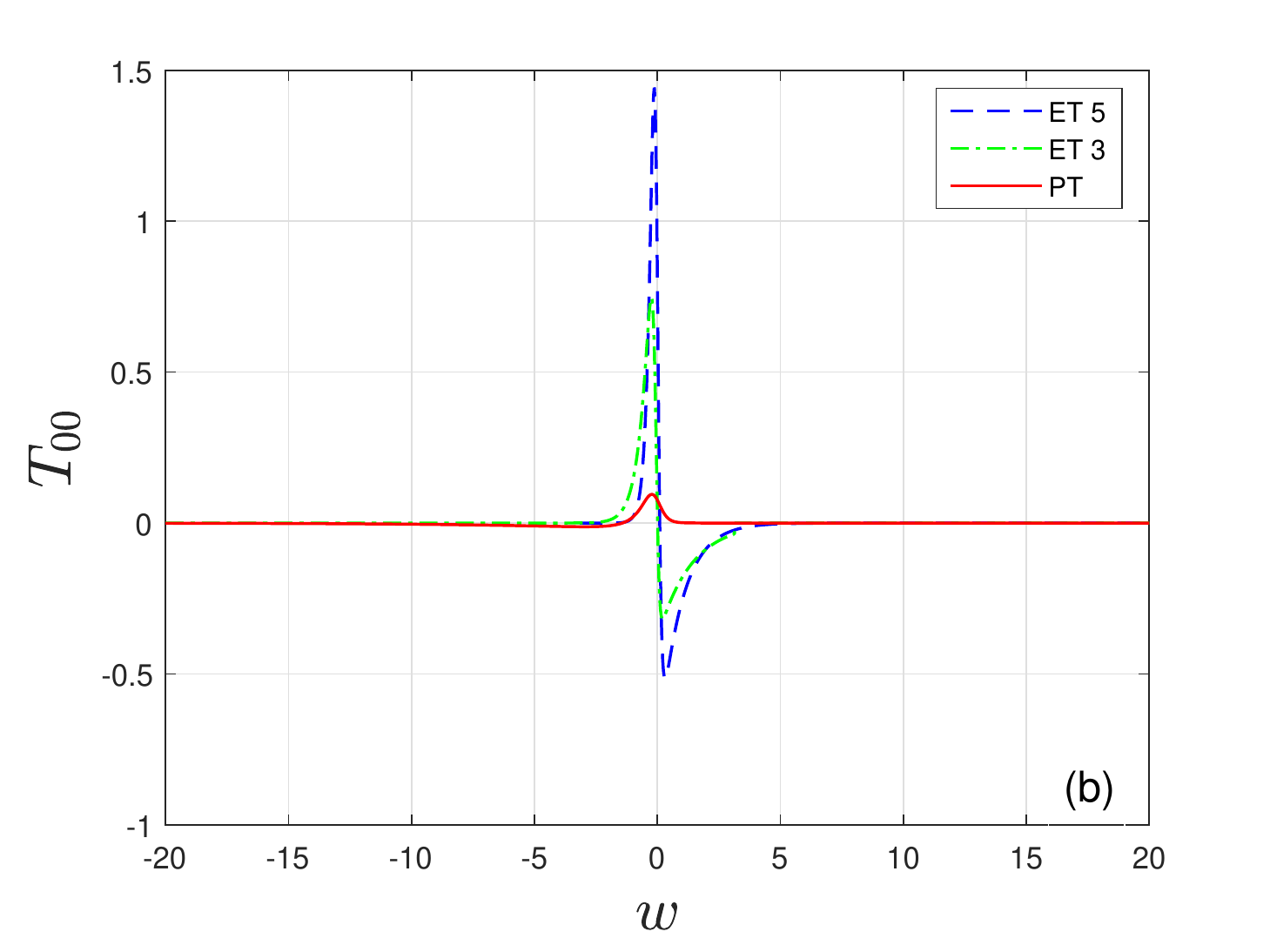}}}
\caption{The energy density $T_{00}$ as a function of the fifth dimension coordinate $w$ for the the ET 5, ET 3 and PT cases for $\lambda=a=1$.
(a) For $n=1$ ($\phi^{10}$), the dashed, dotted-dashed and continuous curves correspond to ET 5 (Eq. (\ref{t1})), ET 3 (Eq. (\ref{t3}))  and PT (Eq. (\ref{t5})), respectively. (b) $n=2$ ($\phi^{18}$). Dashed, dotted-dashed and continuous curves correspond to ET 5 (Eq. (\ref{t2})), ET 3 (Eq. (\ref{t4})) and PT (Eq. (\ref{t6})), respectively.\label{den12}}
\end{figure*}

Figures \ref{r} and \ref{gw}, which represent the Ricci scalar and the mixed Einstein tensor component respectively, show that there is no  $Z_{2}$-symmetry in the bulk on both sides of the brane. Figure \ref{r} shows that the Ricci scalars are singularity-free. Although from Fig. \ref{gw} it is obvious that for all the brane models except the PT model for $\phi^{18}$, the $G^{0}_{0}$ component of the mixed Einstein tensor is zero on the left side and positive on the right side of the brane. Note that in these models the cosmological constant on the brane $G^{A}_{B}\propto\Lambda\delta^{A}_{B}$ 
and consequently the cosmological constant of the bulk completely differs on both sides. In the limit of $w\rightarrow -\infty$ the cosmological constant vanishes, so the bulk is asymptotically Minkowski. While on the other side of the brane, this quantity is a non-zero positive constant which corresponds to a de Sitter spacetime.
Figure \ref{den12} clearly demonstrates that the distribution of energy density for PT branes are more localized and symmetric than for ET ones for both system $\phi^{10}$ and $\phi^{18}$. Besides, as one can see the energy density curve for ET models contain a dip/shoulder on the right side of the brane.

\begin{figure*}[th]
\epsfxsize=9cm\centerline{\hspace{10cm}\epsfbox{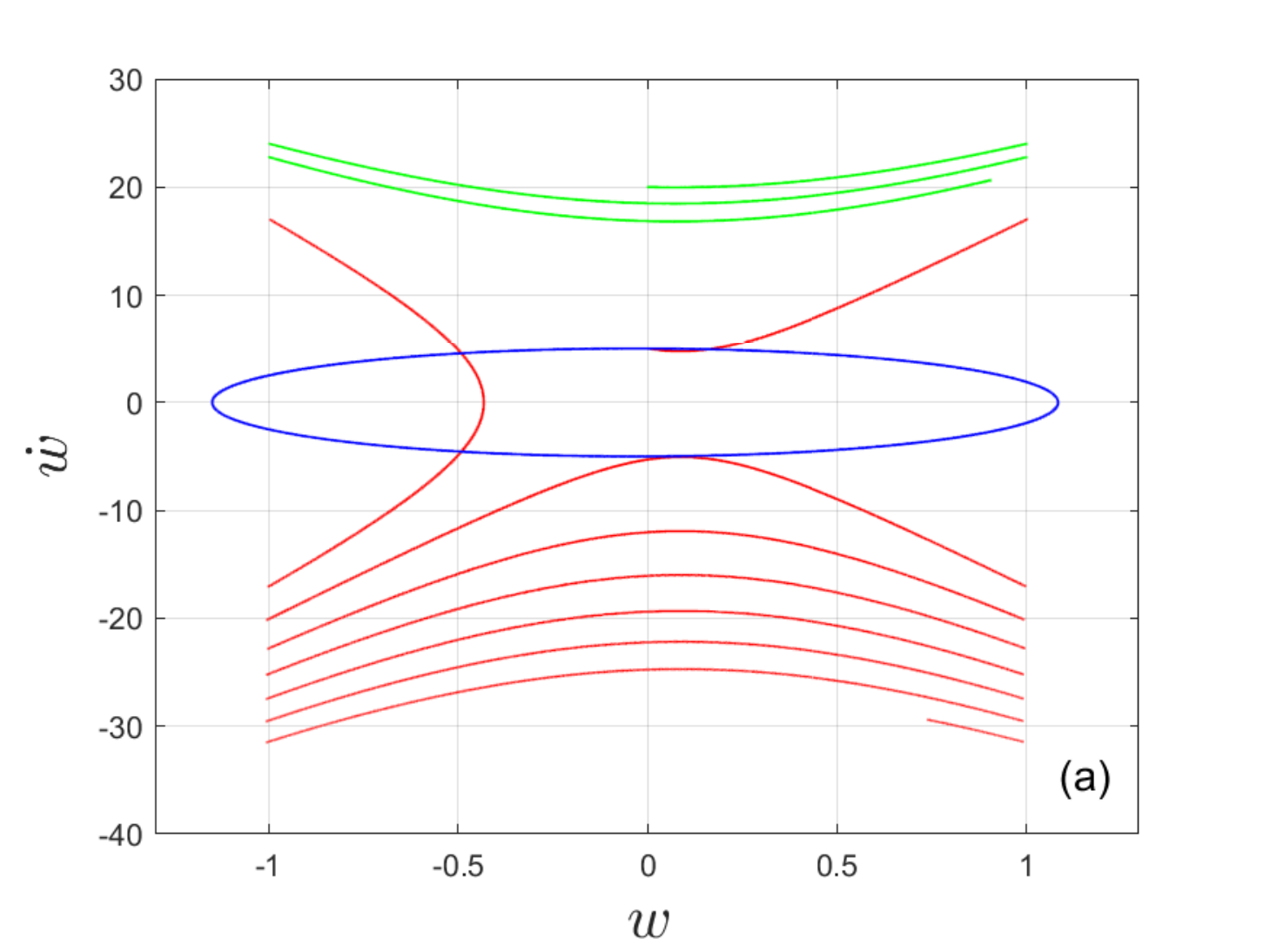}\epsfxsize=9cm\centerline{\hspace{-10cm}\epsfbox{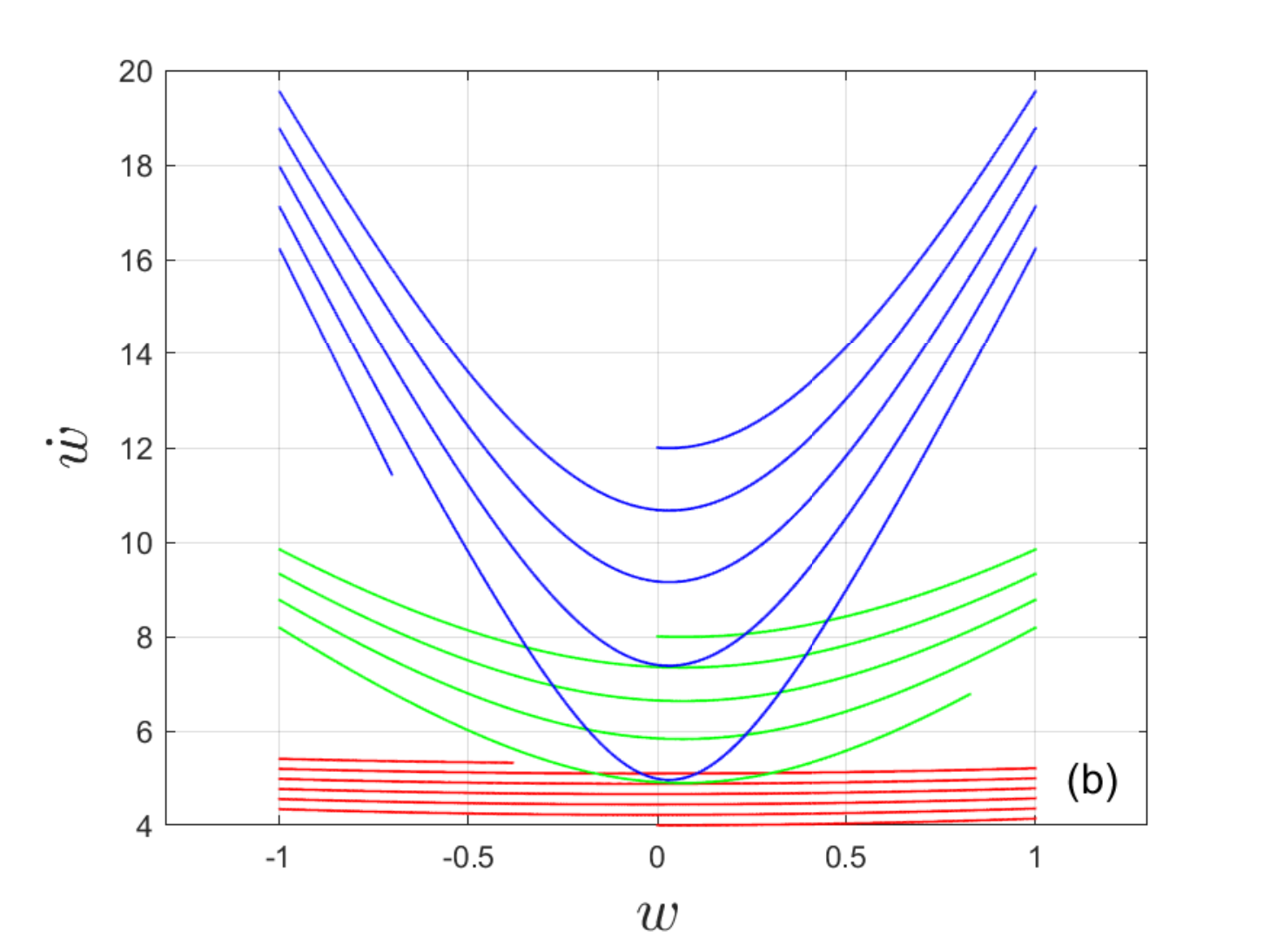}}}
\caption{The phase diagram of the geodesic equation along the fifth dimension $w$ for $\lambda=a=1$; Blue, green and red curves correspond to ET 5, ET 3 and PT models respectively. (a) $\phi^{10}$ Eqs. (\ref{geo1}), (\ref{geo3}) and (\ref{geo5}) (b) $\phi^{18}$ Eqs. (\ref{geo2}), (\ref{geo4}) and (\ref{geo6}) . Closed and open trajectories in this diagram demonstrate the behaviour of confined and unconfined particles near to the brane respectively.\label{ph}}
\end{figure*}

Moreover, it is also interesting to calculate the geodesic equation along the fifth dimension, in order to analyse the particle motion in the neighbourhood of the brane. This investigation is vital to clarify the interaction of material particles to the gravitational field of the brane.
The following phase-space diagram clarifies that among our results of Eqs. (\ref{geo1}), (\ref{geo2}), (\ref{geo3}), (\ref{geo4}), (\ref{geo5}) and (\ref{geo6}) (for $\lambda=1$ and $a=1$), only the $\phi^{10}$ model via ET 5 ( Eq.(\ref{geo1})) supports the confining effects of the particle near the brane (See Fig. \ref{ph}). 

For small oscillations of the amplitude, the relativistic motion reduces to a classical motion in a Newtonian classical potential. As the potential has been considered up to second order, relativistic effects can be ignored and the corresponding quantum energy levels are therefore those of a non-relativistic quantum particle. Thus, if interpreted as a quantum oscillator, one can assign an energy to each quantum state given by $E_n=(n+1/2)\hbar\omega$, where $\hbar \omega=\Omega=\sqrt{F'(w_{0})}$.

\section{Stability}\label{33}

The aim of this section is to study the stability of brane models considered above. In this regard, small perturbations must be applied to the metric and the scalar field via ``axial gauge'' about the static brane, where the resulting equations of motion are linearised  in the transverse and traceless gauge \cite{de,Sa,Af,CM,DA}.

The linearised equation turns out to be a Schr\"{o}dinger-like equation with a potential $U(z)$ which determines the linear modes. 
Unfortunately, for the models considered, this potential is too complicated to be reproduced analytically, or
to be used for finding the corresponding modes. Therefore, we will
only examine the potential near its minimum up to second order in
$z$ \cite{Peyravi:2015bra}. As mentioned above, in order to study the stability of the branes, we choose an ``axial gauge'' in which the metric is perturbed as \cite{de,Sa,Af,CM,DA}:
\begin{equation}
ds^{2}=e^{ 2A}(\eta_{\mu\nu} + \varepsilon h_{\mu\nu} )dx^{\mu} dx^{\nu} - dw^{
2},
\end{equation}
where $g_{\mu\nu}$ represents the four-dimensional AdS or dS metric,
$h_{\mu\nu}$ represents the metric perturbations, and $\varepsilon$ is a
small parameter \cite{Af}. Moreover, in order to render the metric conformally flat, one can choose $dz=e^{-A(w)} dw$. In this case, the corresponding Schr\"{o}dinger equation takes the form \cite{deSouzaDutra:2008gm,de,AB,Sa,Af,DA,CM,BFG}:
\begin{equation}\label{sch}
-\frac{d^{2}\psi(z)}{dz^{2}}+U_{eff}(z)\psi(z)=k^{2}\psi(z),
\end{equation}
with the following effective potential:
\begin{equation}
U_{eff}(z)=-\frac{9}{4}\Lambda+\frac{9}{4}A'^{2}+\frac{3}{2}A''.
\end{equation}
where $\Lambda$ could be positive, negative or zero corresponding to the cosmological constant in the $4D$ spacetime for de Sitter ($dS_{4}$),
anti-de Sitter ($AdS_{4}$) or Minkowski ($M_{4}$), respectively
\cite{Sa,Af}.
In addition to this, by considering the following Hamiltonian corresponding to Eq. (\ref{sch}) \cite{deSouzaDutra:2008gm, AB,Sa,BFG}:
\begin{equation}
H=\left(\frac{d}{dz}+\frac{3}{2}A'(z)\right)\left(-\frac{d}{dz}+\frac{3}{2}A'(z)\right),
\end{equation}
which is obviously Hermitian and therefore leads to real $k$ ($k^{2}\geq0$). One can summarize the above  Schr\"{o}dinger equation (\ref{sch}) as \cite{deSouzaDutra:2008gm}:
\begin{equation}
H\psi =k^2\psi.
\end{equation}
Accordingly, there are no unstable tachyonic excitations in the system
\cite{DA,BFG}.

The solution for the zero modes ($k = 0$) is \cite{DA,YZYX,BFG}:
\begin{equation}
\psi(z)=Ne^{\frac{3A(z)}{2}}  \,,
\end{equation}
where $N$ is a normalization factor\cite{DA,YZYX,BFG} and satisfies:
\begin{eqnarray}
1&=&\int_{-\infty}^{+\infty}dz|\psi_{0}(z)|^{2}=N^{2}\int_{-\infty}^{+\infty}dze^{3A(z)}\nonumber\\
&=&\frac{N^{2}}{l}\int_{-\infty}^{+\infty}dye^{2A(y)}.
\end{eqnarray}
where $y=lw$ is a dimensionless variable.

\begin{figure*}[ht!]
\epsfxsize=9cm\centerline{\hspace{10cm}\epsfbox{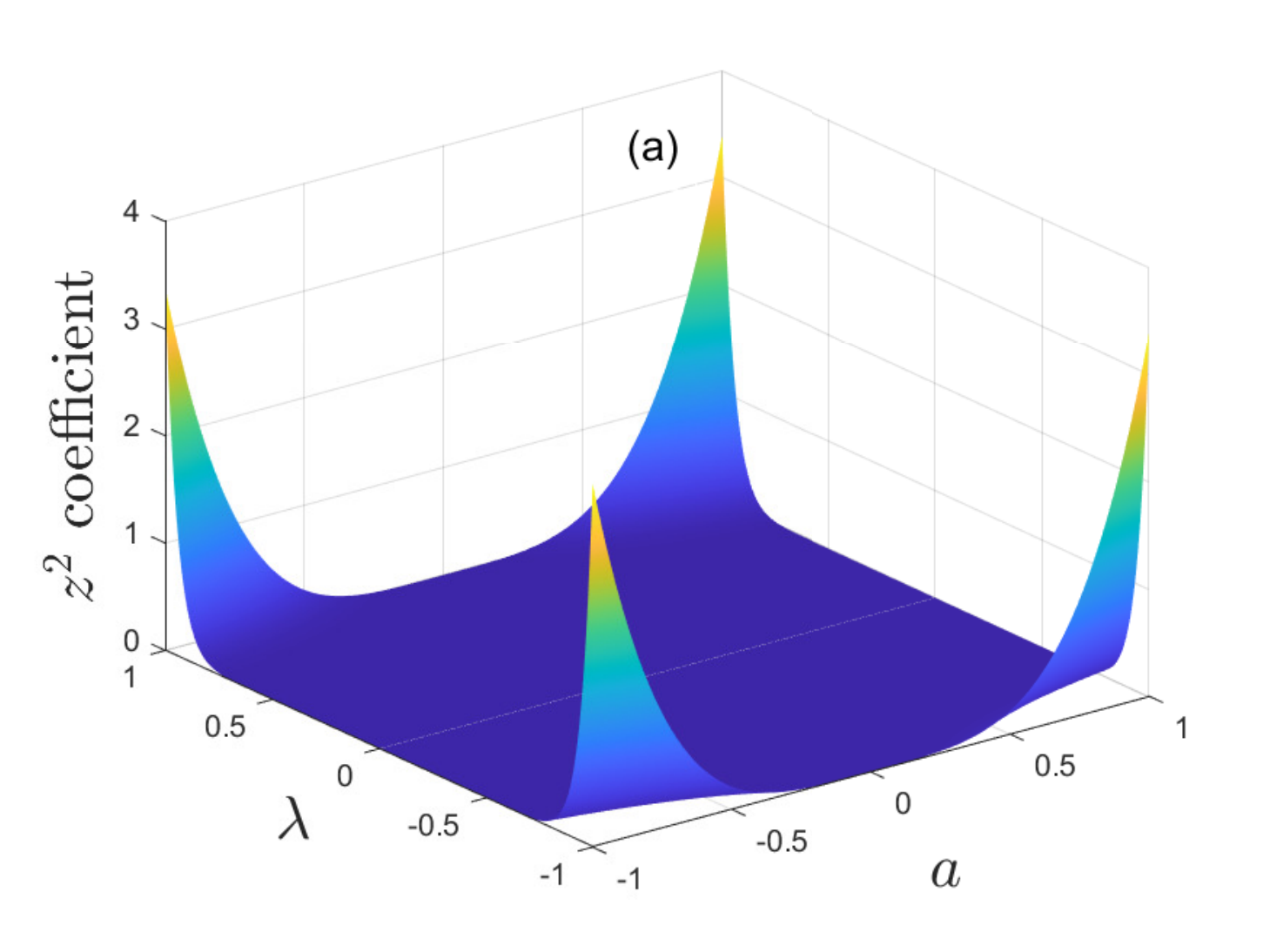}\epsfxsize=9cm\centerline{\hspace{-10cm}\epsfbox{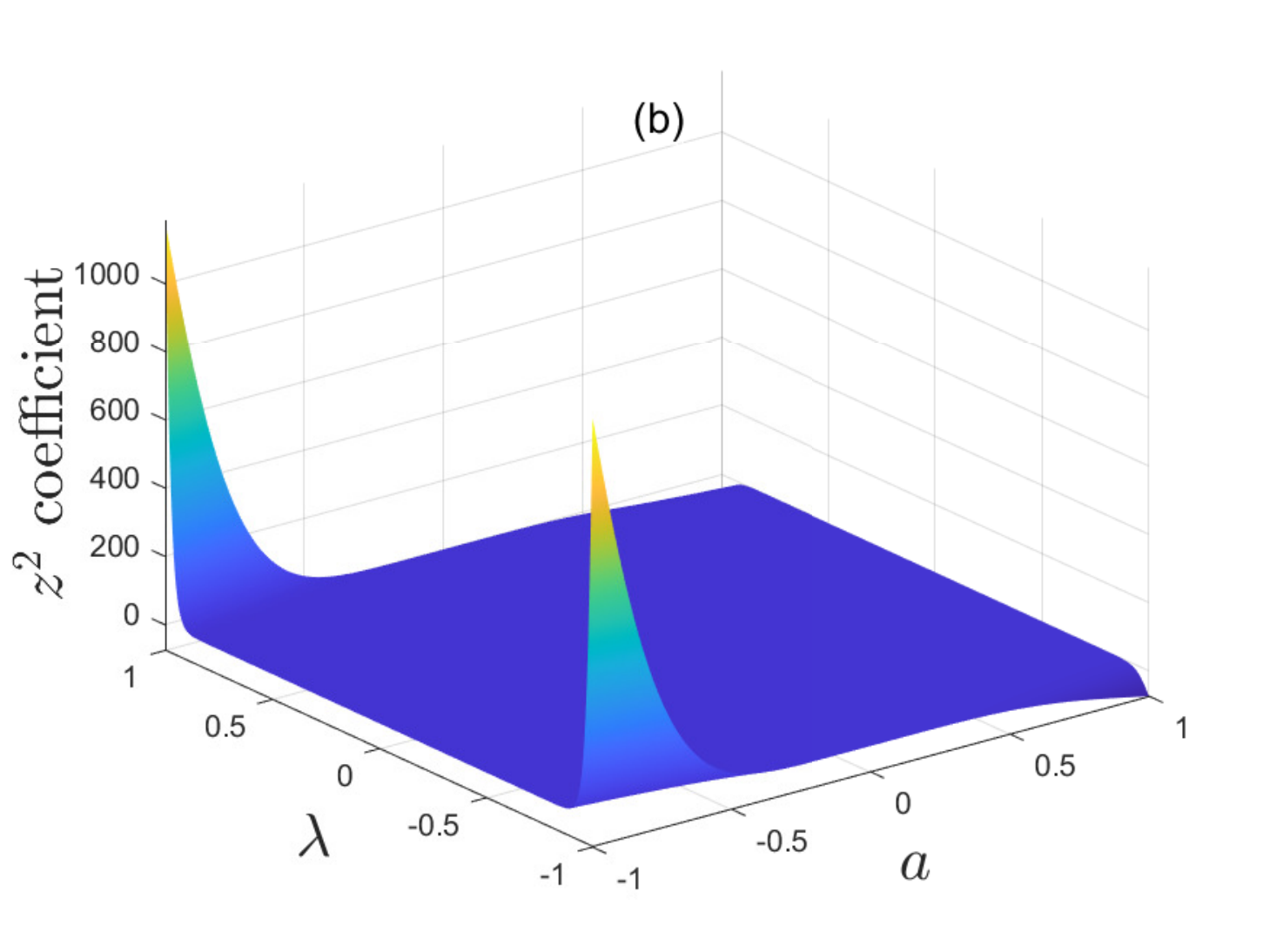}}}
\caption{Stability regions of the potential, where the plots depict the coefficient of the $z^2$ term in the potential of the linearised Schr\"{o}dinger equation as a function of the free parameters
$a$ and $\lambda$ for the  ET 5 (a) $n=1$ ($\phi^{10}$) (b) $n=2$ ($\phi^{18}$)
Note that the sign of the $z^{2}$ term indicates the character of the stability, where the positive sign being stable, while the negative sign indicates instability.}
\label{U1}
\end{figure*}

\begin{figure*}[ht!]
\epsfxsize=9cm\centerline{\hspace{10cm}\epsfbox{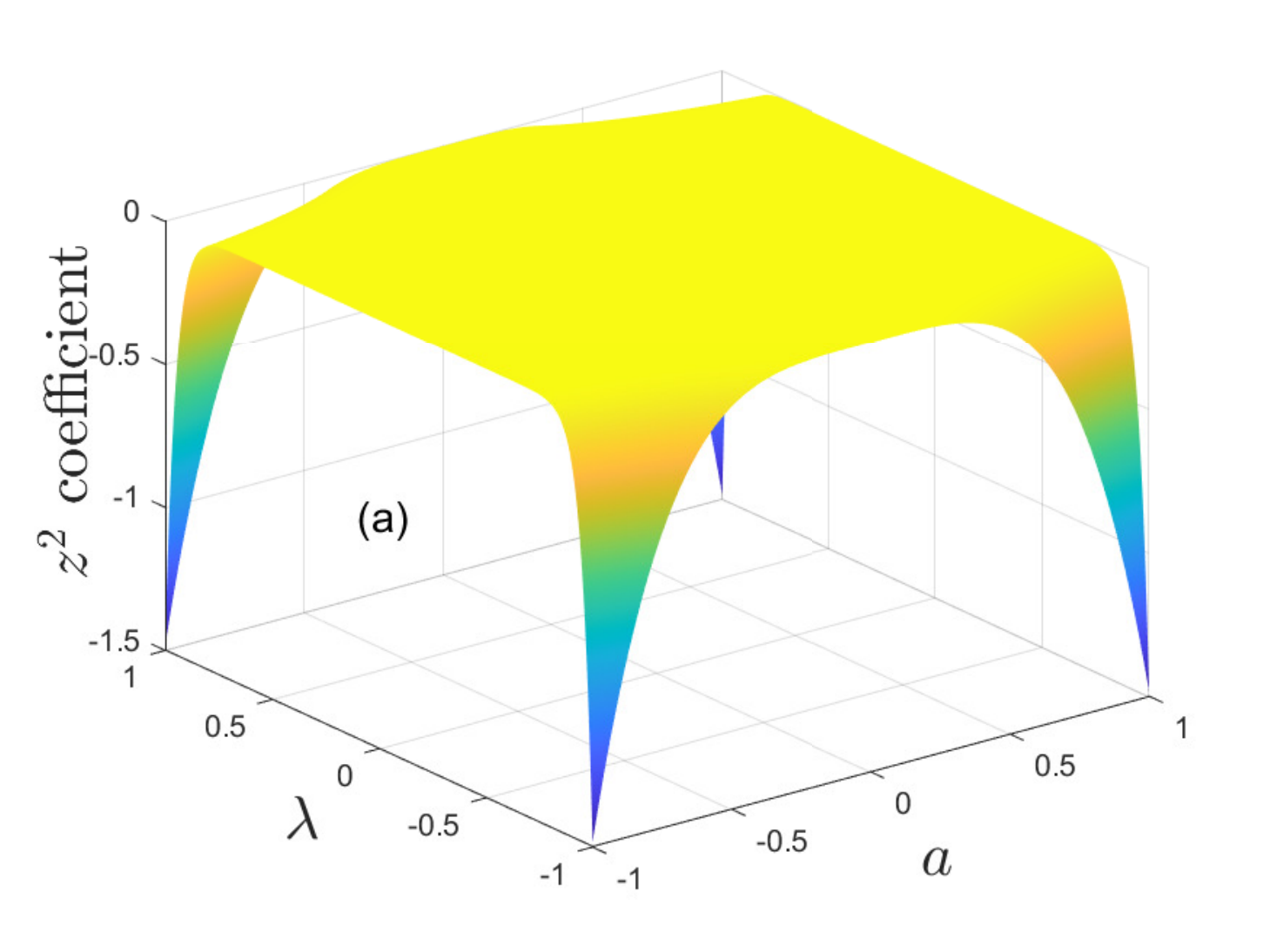}\epsfxsize=9cm\centerline{\hspace{-10cm}\epsfbox{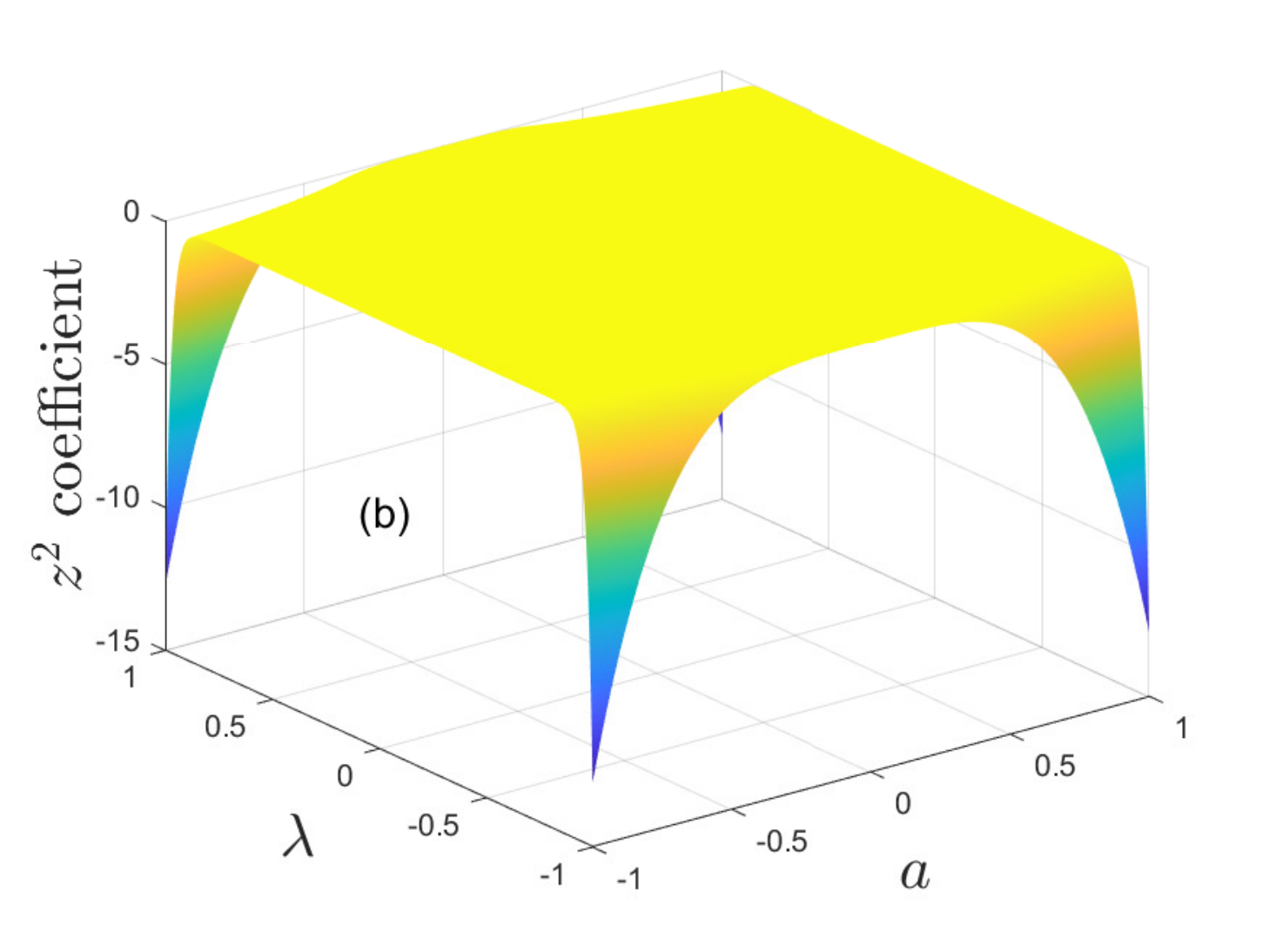}}}
\caption{Stability regions of the potential, where the plots depict the coefficient of the $z^2$ term in the potential of the linearised Schr\"{o}dinger equation as a function of the free parameters
$a$ and $\lambda$ for the ET 3 (a) $n=1$ ($\phi^{10}$) (b) $n=2$ ($\phi^{18}$)
Note that the sign of the $z^{2}$ term indicates the character of the stability, where the positive sign being stable, while the negative sign indicates instability.}
\label{U2}
\end{figure*}

\begin{figure*}[ht!]
\epsfxsize=9cm\centerline{\hspace{10cm}\epsfbox{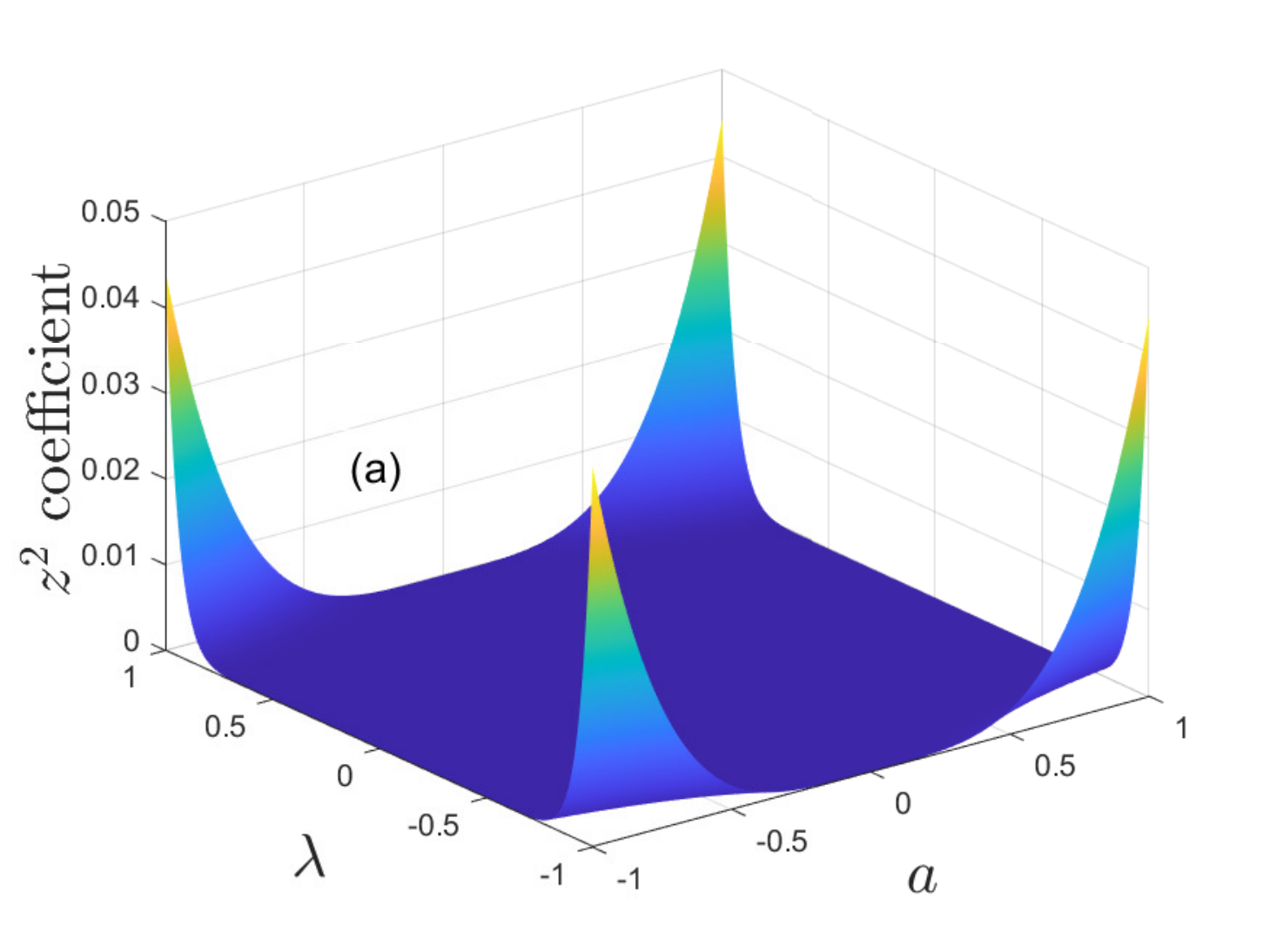}\epsfxsize=9cm\centerline{\hspace{-10cm}\epsfbox{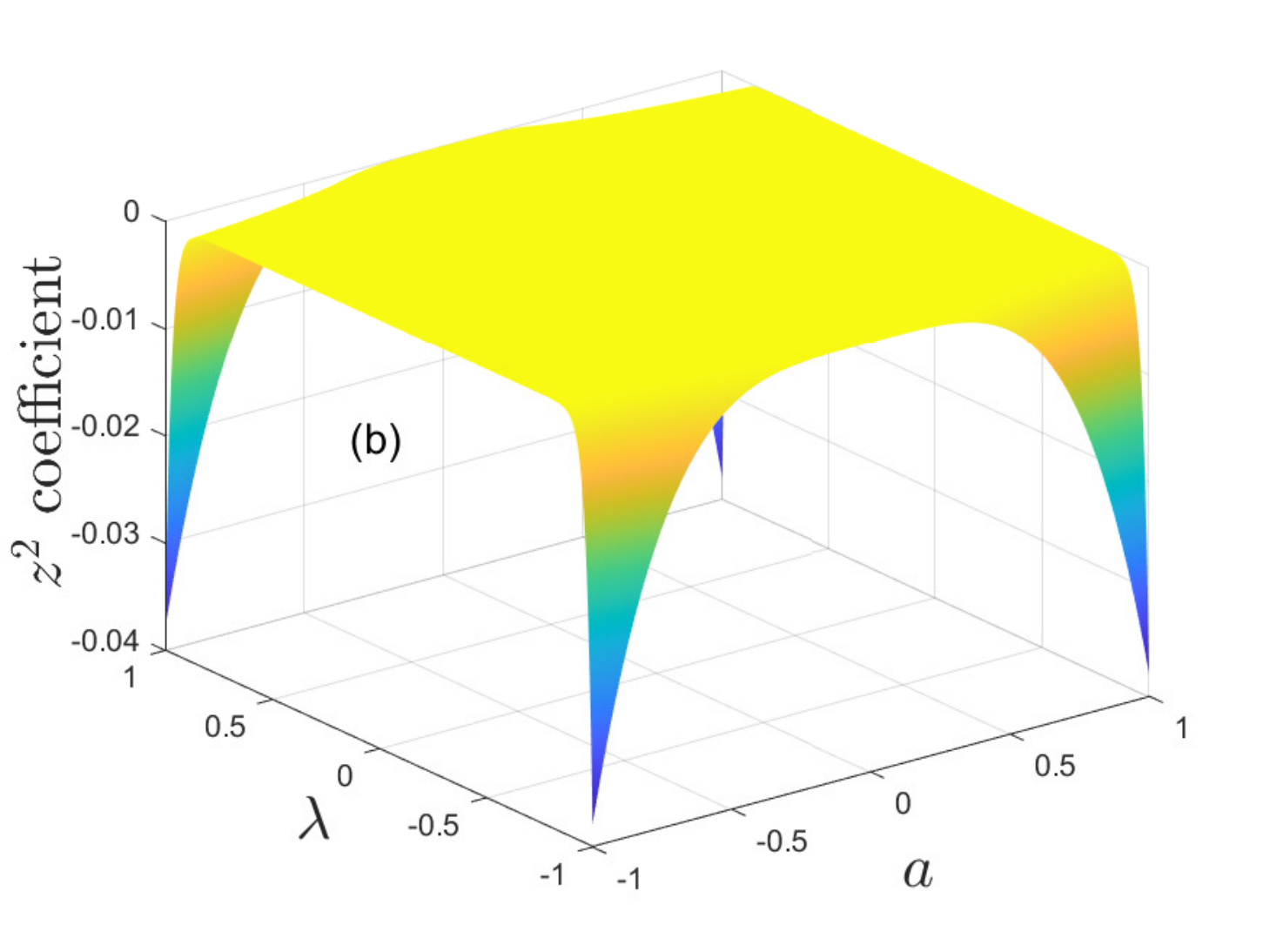}}}
\caption{Stability regions of the potential, where the plots depict the coefficient of the $z^2$ term in the potential of the linearised Schr\"{o}dinger equation as a function of the free parameters
$a$ and $\lambda$ for the PT (a) $n=1$ ($\phi^{10}$) (b) $n=2$ ($\phi^{18}$)
Note that the sign of the $z^{2}$ term indicates the character of the stability, where the positive sign being stable, while the negative sign indicates instability.}
\label{U3}
\end{figure*}

\begin{figure}
\epsfxsize=9cm\centerline{\hspace{10cm}\epsfbox{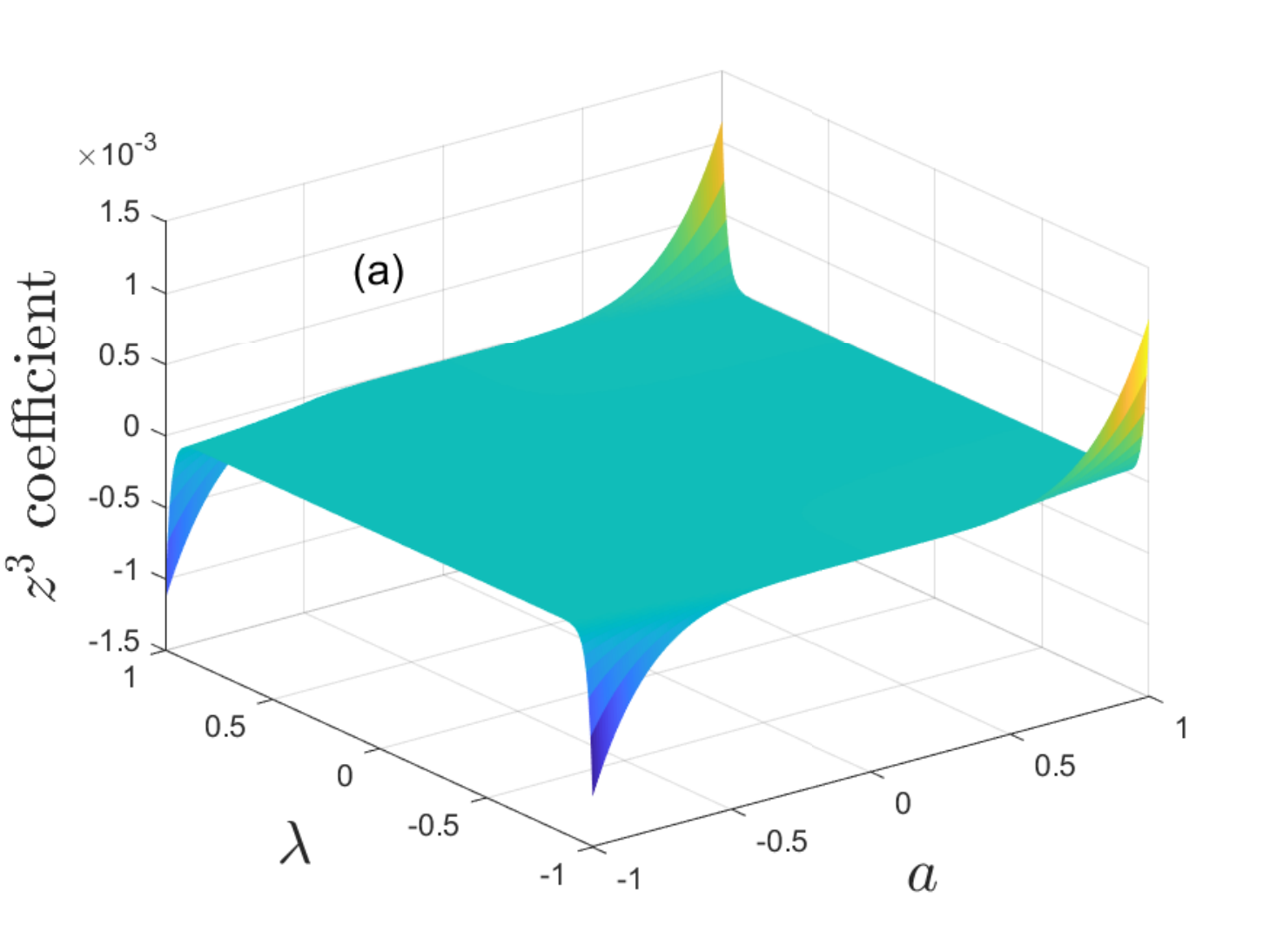}\epsfxsize=9cm\centerline{\hspace{-10cm}\epsfbox{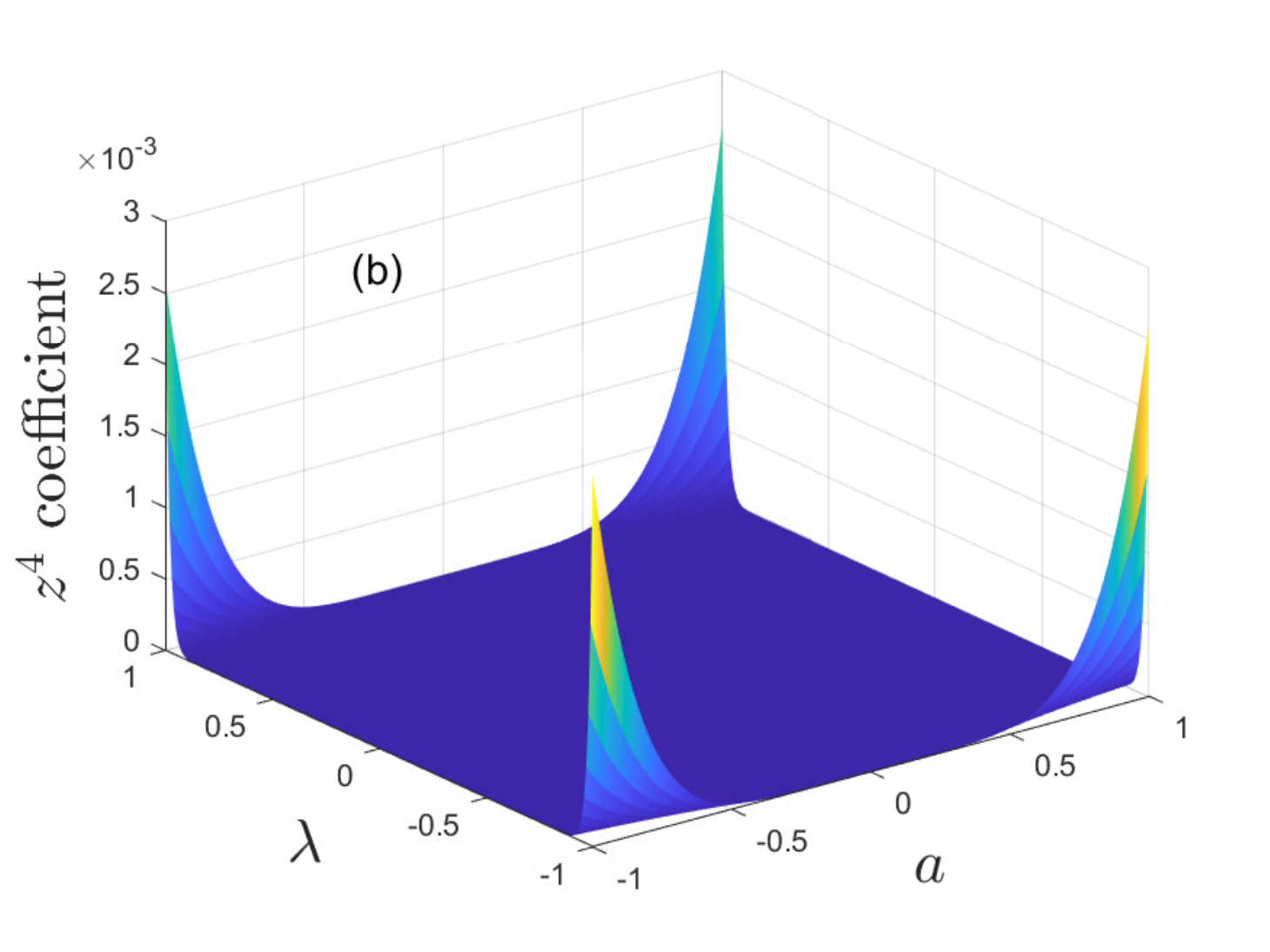}}}
\caption{Stability regions of the potential, where the plots depict the coefficient of the (a) $z^3$ and (b) $z^4$ terms in the potential of the linearised Schr\"{o}dinger equation as a function of the free parameters
$a$ and $\lambda$ for the PT  $n=2$ ($\phi^{18}$).}
\label{U3n}
\end{figure}

On the other hand, as in quantum mechanical systems, we may check for the stability of the
system via the existence of a real frequency, in bound states.
Since the potentials for the three systems have considered in this paper are too complicated to
be solved analytically, we found the corresponding ground
state eigenvalues via expansions in terms of the fifth coordinate $w$.
One can deduce the stability up to $O(w^2)$ by looking at the sign of the
$w^2$ term.
By expanding $U_{eff}(z)$ around $z_{0}$ one obtains:
\begin{equation}\label{com}
U_{eff}(z-z_{0})= U_{0}+(z-z_{0}) U^{\prime}_{0}+\frac{(z-z_{0})^2}{2!}U^{\prime\prime}_{0}+\frac{(z-z_{0})^3}{3!}U^{\prime\prime\prime}_{0}+...+\frac{(z-z_{0})^n}{n!}U^{(n)}_{0}+...
\end{equation}
where 
\begin{equation}
U^{(n)}_{0} =\left(\frac{d^n U}{dz^n}\right)_{z=0}
\end{equation}
Figures \ref{U1} -\ref{U3} display the $z^2$ coefficient of the Taylor expansion of the linearised Schr\"{o}dinger equation potential as a function of the free parameters $a$ and $\lambda$ for the  models considered here. As Figs. \ref{U1}a -\ref{U3}a show, the  ET 5 and PT brane models are stable, while the ET 3 brane involves neutral equilibrium for $n=1$ ($\phi^{10}$).
Furthermore, Figs. \ref{U1}b -\ref{U3}b demonstrate for $n=2$ ($\phi^{18}$), the brane of  ET 5 model is stable, while brane of ET 3 and PT have neutral equilibrium.
In the neutral equilibrium region of the parameters where ($U^{(n)}_{0}=0$), we have to inspect higher order terms in Eq. (\ref{com}). As one can find from Fig. \ref{U3n}, the coefficients  $z^3$ and  $z^4$ terms for the $\phi^{18}$ model are also zero, which guarantee the neutral stability region of the brane in Fig. \ref{U3}.

\section{Conclusion}\label{44}

In this work, we have investigated higher order field theory kinks and obtained exact thick brane models inspired by three different potentials for $\phi^{10}$ and $\phi^{18}$. The confining effect of the scalar field in all of these models were confirmed by examining
the geodesic equation for a test particle moving normal to the brane. These models are particularly  interesting, since they have more than two pairs of solitions and anti-solitons which live in
different topological sectors. In particular, we have considered several solutions with power-law (PT) and exponential tails (ET). We have shown that the resulting brane of the ET 5 and ET 3 specific models do not have $Z_2$-symmetry, in general, where the center of the brane may be displaced from $w=0$ and the potential will not be an odd function of $w$ in general.  However, by decreasing the $\lambda$ parameter it is possible to make the vacua of the effective potential degenerate, in which case the $Z_2$-symmetry is restored.

To study the brane scenarios considered in this work, we have summarizes the results in the figures presented, and presented all the relevant quantities, namely, the auxiliary function, $W$, the modified potential, $V$, the warp factor, $A$, Ricci scalar $R$, the mixed Einstein tensor components $G_{D}^{C}$, the energy momentum tensor $T_{00}$ and geodesic equation in the Appendixes \ref{appendix:AA}-\ref{appendix:CC}, due to their lengthy and messy characters. 
Finally, we examined the stability of the thick branes, by determining the sign of the $w^2$ term in the expansion of the potential for the resulting Schr\"{o}dinger-like
equation. It turns out that two of the three models of $\phi^{10}$ brane are stable, while
another contains unstable modes for certain ranges of the model parameters. However, for the $\phi^{18}$ branes specific solutions are stable, while the others involve neutral equilibrium.
As future work, an interesting line of investigation would be the study of explicit kink solutions by considering the potential in the other sectors.

\acknowledgments{M.P. acknowledges the support of
Ferdowsi University of Mashhad and S.N. acknowledges the support of
University of Neyshabur.
FSNL acknowledges support from the Funda\c{c}\~{a}o para a Ci\^{e}ncia e a Tecnologia (FCT) Scientific Employment Stimulus contract with reference CEECINST/00032/2018, and funding through the research grants UIDB/04434/2020, UIDP/04434/2020, PTDC/FIS-OUT/29048/2017, CERN/FIS-PAR/0037/2019 and PTDC/FIS-AST/0054/2021.
}




\appendix
\numberwithin{equation}{section}
\renewcommand{\theequation}{\thesection\arabic{equation}}


\begin{widetext}

\section{Properties of Thick Brane with Exponential Tail via $5$ degenerate minima potential (ET 5)}\label{appendix:AA}
Due to their extremely lengthy and messy character,  
the auxiliary function $W$, modified potential, warp factor, energy momentum tensor, geodesic equation, Ricci and the mixed Einstein tensor components of the considered models are given by following subsections respectively (we have considered that $\mu = (0,1,2,3)$ and $c_1$ is an integration constant).

\subsection{ Case $n=1$ ($\phi^{10}$)}

\subsubsection{Auxiliary function and warp factor}

The auxiliary function $W$ for this case is given by:
\begin{equation}\label{W5}
W(\phi)=\frac{ S_{1}}{24} \phi ^2 \left(\frac{2 \phi ^4}{a ^4}-\frac{9 \phi ^2}{a ^2}+12\right)\,,
\end{equation}
which provides the following expression for the potential $V(\phi)$:
\begin{equation}\label{V5}
V(\phi)=\frac{ S_{1}^2}{32} \left[\phi ^2 \left(\frac{\phi ^4}{a^4}-\frac{3 \phi ^2}{a ^2}+2\right)^2-\frac{1}{54} \phi ^4 \left(\frac{2 \phi ^4}{a ^4}-\frac{9 \phi ^2}{a ^2}+12\right)^2\right]\,.
\end{equation}

The warp factor $A$ takes the form:
\begin{align}\label{Ap1}
e^{2A}=e^{\frac{ a ^2}{9} \left(\frac{1}{\sqrt{e^{S_{1} w}+1}}-1\right)} \left[\frac{a^6}{\left(e^{S_{1} w}+1\right)^{3/2} \left(\sqrt{e^{S_{1} w}+1}+1\right)^2}\right]^{\frac{a ^2}{18}} \,.
\end{align}

\subsubsection{Ricci scalar}

The Ricci scalar, $R$, for this model is given by:
\begin{multline}\label{Ap2}
R=-\frac{a^2 S_{1}^2 e^{S_{1} w}}{1296 \left(e^{S_{1} w}+1\right)^3 \left(\sqrt{e^{S_{1} w}+1}+1\right)^2} \times \\
\times \left\{125 a^2 e^{3 S_{1} w}+e^{2 S_{1} w} \left[25 a^2 \left(10 \sqrt{e^{S_{1} w}+1}+19\right)-432\right]+2 e^{S_{1} w} \left[5 a^2 \left(35 \sqrt{e^{S_{1} w}+1}+37\right)\right.\right.\\
\left.\left. -216 \left(3 \sqrt{e^{S_{1} w}+1}+5\right)\right]-1728 \left(\sqrt{e^{S_{1} w}+1}+1\right)\right\}.
\end{multline}
\begin{align}\label{Ap2l}
\lim_{w\rightarrow+\infty} R&=-\frac{125 a^4S_{1}^2}{1296},\nonumber\\
\lim_{w\rightarrow-\infty} R&=0,\nonumber\\
\lim_{w\rightarrow0} R&=-\frac{a^2 \left\{25 \left(10 \sqrt{2}+19\right) a^2+125 a^2+2 \left[5 \left(35 \sqrt{2}+37\right) a^2-216 \left(3 \sqrt{2}+5\right)\right]-1728 \left(\sqrt{2}+1\right)-432\right\} S_{1}^2}{10368 \left(\sqrt{2}+1\right)^2}\,.
\end{align}

\subsubsection{Einstein tensor components}

The mixed Einstein tensor components $G_{D}^{C}$, are given by the following equations, respectively:
\begin{eqnarray} \label{Ap3}
G^{\mu}_{\mu}&=& \frac{a^2 S_{1}^2 e^{S_{1} w}}{864 \left(e^{S_{1} w}+1\right)^3 \left(\sqrt{e^{S_{1} w}+1}+1\right)^2} \times \nonumber\\
&&\times \left\{e^{S_{1} w} \left[a^2 \left(70 \sqrt{e^{S_{1} w}+1}+5 e^{S_{1} w} \left(10 \sqrt{e^{S_{1} w}+1}+5 e^{S_{1} w}+19\right)+74\right)\right.\right.\nonumber\\
&&\left.\left. -108 \left(3 \sqrt{e^{S_{1} w}+1}+e^{S_{1} w}+5\right)\right] -432 \left(\sqrt{e^{S_{1} w}+1}+1\right)\right\}\,,
\end{eqnarray}
\begin{align}\label{Ap3l}
\lim_{w\rightarrow0}G^{\mu}_{\mu}=\frac{a^2 \left\{\left[70 \sqrt{2}+5 \left(10 \sqrt{2}+24\right)+74\right] a^2-432 \left(\sqrt{2}+1\right)-108 \left(3 \sqrt{2}+6\right)\right\} S_{1}^2}{6912 \left(\sqrt{2}+1\right)^2}\,,
\end{align}
\begin{align} \label{Ap4}
G^{4}_{4}=\frac{a^4 S_{1}^2 e^{2 S_{1} w} \left[70 \sqrt{e^{S_{1} w}+1}+5 e^{S_{1} w} \left(10 \sqrt{e^{S_{1} w}+1}+5 e^{S_{1} w}+19\right)+74\right]}{864 \left(e^{S_{1} w}+1\right)^3 \left(\sqrt{e^{S_{1} w}+1}+1\right)^2}\,,
\end{align}
\begin{align}\label{Ap34l}
\lim_{w\rightarrow+\infty} G_{D}^{C}&=\frac{25 a^4 S_{1}^2}{864},\nonumber\\
\lim_{w\rightarrow-\infty}G_{D}^{C}&=0 \,.
\end{align}

\subsubsection{Energy momentum tensor}

The energy momentum tensor $T_{00}$:
\begin{multline}\label{t1}
T_{00}= \frac{1}{1728} \left[a ^2 S_{1}^2 e^{\frac{a ^2}{9}  \left(\frac{1}{\sqrt{e^{S_{1} w}+1}}-1\right)} \left(\frac{a ^6}{\left(e^{S_{1} w}+1\right)^{3/2} \left(\sqrt{e^{S_{1} w}+1}+1\right)^2}\right)^{\frac{a ^2}{18}} \right] \times\\
\times \left[a ^2 \left(-\frac{9}{\left(e^{S_{1} w}+1\right)^2}-\frac{4}{\left(e^{S_{1} w}+1\right)^3}+\frac{20}{\left(e^{S_{1} w}+1\right)^{3/2}}-\frac{12}{\left(e^{S_{1} w}+1\right)^{5/2}}+\frac{30}{e^{S_{1} w}+1}-25\right)
\right.\\
\left.
+\frac{108 e^{S_{1} w} \left(\sqrt{e^{S_{1} w}+1}+1\right)}{\left(e^{S_{1} w}+1\right)^{5/2}}\right] \,.
\end{multline}

\subsubsection{Geodesic equation}

The geodesic equation is given by:
\begin{multline}\label{geo1}
\ddot{w}+c_{1}^{2}\frac{ 2^{\frac{1}{12} \left(a ^2-90\right)}}{81} \left(2 \sqrt{2}+3\right)^{\frac{a ^2}{18}-1} e^{-\frac{1}{18} \left(\sqrt{2}-2\right) a ^2} \left(\frac{1}{a ^6}\right)^{\frac{a ^2}{18}} a ^2 \left[\left(97 \sqrt{2}+120\right) a ^2+108 \left(5 \sqrt{2}+7\right)\right]  S_{1}^2 w
	\\
\approx c_{1}^{2}\frac{1}{9} \left(\sqrt{2}-7\right) 2^{\frac{a ^2}{12}-4} \left(2 \sqrt{2}+3\right)^{\frac{a ^2}{18}} e^{-\frac{1}{18} \left(\sqrt{2}-2\right) a ^2} \left(\frac{1}{a ^6}\right)^{\frac{a ^2}{18}} a ^2 S_{1}.
\end{multline}

\subsection{ Case $n=2$ ($\phi^{18}$)}

\subsubsection{Auxiliary function and warp factor}

The auxiliary function $W$ for this case is given by:
\begin{equation}\label{W52}
W(\phi)=\frac{S_{2}}{40} \phi ^2 \left(\frac{\phi ^8}{a ^8}-\frac{5 \phi ^4}{a ^4}+10\right)\,,
\end{equation}
which provides the following expression for the potential $V(\phi)$:
\begin{equation}\label{V52}
V(\phi)=\frac{S_{2}^2}{128}  \left[\phi ^2 \left(\frac{\phi ^8}{a ^8}-\frac{3 \phi ^4}{a ^4}+2\right)^2-\frac{2}{75} \phi ^4 \left(\frac{\phi ^8}{a ^8}-\frac{5 \phi ^4}{a ^4}+10\right)^2\right]\,,
\end{equation}

The warp factor $A$ takes the form:
\begin{multline}\label{Ap5}
e^{2A}=\exp \left\{\frac{1}{15} \left[2 a ^2 \left(\sqrt{2} \tanh ^{-1}\left(\frac{\sqrt{1-\frac{1}{\sqrt{e^{S_{2} w}+1}}}}{\sqrt{2}}\right)-3 \tanh ^{-1}\left(\sqrt{1-\frac{1}{\sqrt{e^{S_{2} w}+1}}}\right)\right)-a ^2 \sqrt{1-\frac{1}{\sqrt{e^{S_{2} w}+1}}}\right]\right\} \,.
\end{multline}

\subsubsection{Ricci scalar}

The Ricci scalar, $R$, for this model is given by:
\begin{eqnarray}\label{Ap6}
R&=&\frac{a^2 S_{2}^2 e^{2  S_{2} w}}{720 \left(e^{ S_{2} w}+1\right)^{5/2} \sqrt{1-\frac{1}{\sqrt{e^{ S_{2} w}+1}}} \left(\sqrt{e^{ S_{2} w}+1}-1\right) \left(\sqrt{e^{ S_{2} w}+1}+1\right)^2 \left(\sqrt{e^{ S_{2} w}+1}-e^{ S_{2} w}-1\right)} \times
	\nonumber\\
&&\times \left\{36 a^2 e^{3  S_{2} w} \sqrt{\frac{-\sqrt{e^{ S_{2} w}+1}+e^{ S_{2} w}+1}{e^{ S_{2} w}+1}}-16 a^2 \sqrt{\frac{-\sqrt{e^{ S_{2} w}+1}
 +e^{ S_{2} w}+1}{e^{ S_{2} w}+1}} \left(\sqrt{e^{ S_{2} w}+1}-1\right)\right.\nonumber\\
&&\left. +e^{2  S_{2} w} \left(93 a^2 \sqrt{\frac{-\sqrt{e^{ S_{2} w}+1}+e^{ S_{2} w}+1}{e^{ S_{2} w}+1}}-60\right)\right.\nonumber\\
&&\left.\left. -e^{ S_{2} w} \left[a^2 \sqrt{\frac{-\sqrt{e^{ S_{2} w}+1}+e^{ S_{2} w}+1}{e^{ S_{2} w}+1}} \left(15 \sqrt{e^{ S_{2} w}+1}-73\right) +60 \left(\sqrt{e^{ S_{2} w}+1}+1\right)\right]\right\}\right. \,,
\end{eqnarray}
\begin{align}\label{Ap6l}
\lim_{w\rightarrow+\infty} R&=-\frac{ a^4 S_{2}^2}{20}\,,\nonumber\\
\lim_{w\rightarrow-\infty} R&=0\,,\nonumber\\
\lim_{w\rightarrow0} R&=-\frac{a^2 \left[-16 \sqrt{1-\frac{1}{\sqrt{2}}} \left(\sqrt{2}-1\right) a^2-\sqrt{1-\frac{1}{\sqrt{2}}} \left(15 \sqrt{2}-73\right) a^2+129 \sqrt{1-\frac{1}{\sqrt{2}}} a^2-60 \left(\sqrt{2}+1\right)-60\right] S_{2}^2}{5760 \sqrt{1-\frac{1}{\sqrt{2}}}} \,.
\end{align}

\subsubsection{Einstein tensor components}

The mixed Einstein tensor components $G_{D}^{C}$:
\begin{eqnarray}\label{Ap7}
G^{\mu}_{\mu}&=&-\frac{a^2  S_{2}^2 }{2400 \left(e^{ S_{2} w}+1\right)^3 \sqrt{1-\frac{1}{\sqrt{e^{ S_{2} w}+1}}}}
\Bigg\{-36 a^2 e^{3  S_{2} w} \sqrt{1-\frac{1}{\sqrt{e^{ S_{2} w}+1}}}+16 a^2 \left(\sqrt{e^{ S_{2} w}+1}-1\right) \sqrt{1-\frac{1}{\sqrt{e^{ S_{2} w}+1}}}\nonumber\\
&& +e^{ S_{2} w} \left[a^2 \left(15 \sqrt{e^{ S_{2} w}+1}-73\right) \sqrt{1-\frac{1}{\sqrt{e^{ S_{2} w}+1}}}+75 \left(\sqrt{e^{ S_{2} w}+1}+1\right)\right]
	\nonumber \\
&&+e^{2  S_{2} w} \left(75-93 a^2 \sqrt{1-\frac{1}{\sqrt{e^{ S_{2} w}+1}}}\right)\Bigg\} \,,
\end{eqnarray}
\begin{align}\label{Ap7l}
\lim_{w\rightarrow0}G^{\mu}_{\mu}=\frac{a^2 \left[16 \left(\sqrt{2}-1\right) \sqrt{1-\frac{1}{\sqrt{2}}} a^2+\left(15 \sqrt{2}-73\right) \sqrt{1-\frac{1}{\sqrt{2}}} a^2-129 \sqrt{1-\frac{1}{\sqrt{2}}} a^2+75 \left(\sqrt{2}+1\right)+75\right]  S_{2}^2}{19200 \sqrt{1-\frac{1}{\sqrt{2}}}}\,,
\end{align}
\begin{align}\label{Ap8}
G^{4}_{4}=\frac{a^4  S_{2}^2 \left[e^{ S_{2} w} \left(-15 \sqrt{e^{ S_{2} w}+1}+93 e^{ S_{2} w}+36 e^{2  S_{2} w}+73\right)-16 \left(\sqrt{e^{ S_{2} w}+1}-1\right)\right]}{2400 \left(e^{ S_{2} w}+1\right)^3} \,,
\end{align}
\begin{align}\label{Ap78l}
\lim_{w\rightarrow+\infty} G_{D}^{C}&=\frac{3 a^4 S_{2}^2}{200},\nonumber\\
\lim_{w\rightarrow-\infty} G_{D}^{C}&=0 \,.
\end{align}

\subsubsection{Energy momentum tensor}

The energy momentum tensor $T_{00}$ is given by:
\begin{multline}\label{t2}
T_{00}=\frac{a ^2  S_{2}^2}{9600 \left(e^{ S_{2} w}+1\right)^{5/2}} \\
\exp \left\{-\frac{ a ^2}{15} \left[\sqrt{1-\frac{1}{\sqrt{e^{ S_{2} w}+1}}}+6 \tanh ^{-1}\left(\sqrt{1-\frac{1}{\sqrt{e^{ S_{2} w}+1}}}\right)-2 \sqrt{2} \tanh ^{-1}\left(\frac{\sqrt{1-\frac{1}{\sqrt{e^{ S_{2} w}+1}}}}{\sqrt{2}}\right)\right]\right\} \times \\
\times \left[-32 a ^2 \left(\sqrt{e^{ S_{2} w}+1}-1\right)-6 a ^2 e^{ S_{2} w} \left(19 \sqrt{e^{ S_{2} w}+1}-5\right)+e^{2  S_{2} w} \left(\frac{150 \sqrt[4]{e^{ S_{2} w}+1}}{\left(\sqrt{e^{ S_{2} w}+1}-1\right)^{3/2}}-72 a ^2 \sqrt{e^{ S_{2} w}+1}\right)\right]\,.
\end{multline}

\subsubsection{Geodesic equation}

The geodesic equation is given by:
\begin{multline}\label{geo2}
\ddot{w}+c_{1}^{2}\frac{\left[\left(140 \sqrt{2}-249\right) a ^2-75 \sqrt{2-\sqrt{2}} \left(\sqrt{2}+1\right)\right] a ^2  S_{2}^2  }{28800 \left(\sqrt{2}-2\right)}\\
\exp \left[\frac{ a ^2}{30} \left(\sqrt{4-2 \sqrt{2}}+12 \tanh ^{-1}\left(\sqrt{1-\frac{1}{\sqrt{2}}}\right)-4 \sqrt{2} \tanh ^{-1}\left(\frac{\sqrt{2-\sqrt{2}}}{2}\right)\right)\right] w\\
\approx -c_{1}^{2}\frac{\left(3 \sqrt{2}+13\right) }{240} \sqrt{1-\frac{1}{\sqrt{2}}} a ^2  S_{2} \exp \left[\frac{1}{30} a ^2 \left(\sqrt{4-2 \sqrt{2}}+12 \tanh ^{-1}\left(\sqrt{1-\frac{1}{\sqrt{2}}}\right)-4 \sqrt{2} \tanh ^{-1}\left(\frac{\sqrt{2-\sqrt{2}}}{2}\right)\right)\right] \,.
\end{multline}

\section{Properties of Thick Brane with Exponential Tail via $3$ degenerate minima potential (ET 3)}\label{appendix:BB}

\subsection{ Case $n=1$ ($\phi^{10}$)}

\subsubsection{Auxiliary function and warp factor}

The auxiliary function $W$ for this case is given by:
\begin{equation}\label{w3}
W(\phi)=\frac{ S_{1}}{12} \phi ^2 \left(3-\frac{\phi ^4}{a ^4}\right)\,,
\end{equation}
which provides the following expression for the potential $V(\phi)$:
\begin{equation}\label{V3}
V(\phi)=\frac{-S_{1}^2}{864} \phi ^2 \left(\frac{2 \phi ^{10}}{a ^8}-\frac{27 \phi ^8}{a ^8}-\frac{12 \phi ^6}{a ^4}+\frac{54 \phi ^4}{a ^4}+18 \phi ^2-27\right) \,.
\end{equation}

The warp factor $A$ takes the form:
\begin{align}\label{Ap9}
e^{2A}=\exp \left[\frac{ -a ^2}{9} \left(\frac{1}{\sqrt{e^{S_{1} (-w)}+1}}+2 \tanh ^{-1}\left(\frac{1}{\sqrt{e^{S_{1} (-w)}+1}}\right)\right)\right] \,.
\end{align}

\subsubsection{Ricci scalar}

The Ricci scalar, $R$, for this model is given by:
\begin{align}\label{Ap10}
R=-\frac{a^2 S_{1}^2 e^{S_{1} w} \left[5 a^2 \left(2 e^{S_{1} w}+3\right)^2-108 \sqrt{e^{-S_{1} w}+1}\right]}{324 \left(e^{S_{1} w}+1\right)^3} \,,
\end{align}
\begin{align}\label{Ap9l}
\lim_{w\rightarrow+\infty} R&=-\frac{5 a^4S_{1}^2}{81} \,,\nonumber\\
\lim_{w\rightarrow-\infty} R&=0 \,,\nonumber\\
\lim_{w\rightarrow0} R&=-\frac{a^2 \left(125 a^2-108 \sqrt{2}\right) S_{1}^2}{2592} \,.
\end{align}

\subsubsection{Einstein tensor components}

The mixed Einstein tensor components $G_{D}^{C}$:
\begin{align}\label{Ap11}
G^{\mu}_{\mu}=\frac{a^2 S_{1}^2 e^{S_{1} w} \left[a^2 \left(2 e^{S_{1} w}+3\right)^2-27 \sqrt{e^{-S_{1} w}+1}\right]}{216 \left(e^{S_{1} w}+1\right)^3} \,,
\end{align}
\begin{align}\label{Ap11l}
\lim_{w\rightarrow0}G^{\mu}_{\mu}&=\frac{a^2 \left(25 a^2-27 \sqrt{2}\right)S_{1}^2}{1728} \,,
\end{align}
\begin{align}\label{Ap12}
G^{4}_{4}=\frac{a^4 S_{1}^2 e^{S_{1} w} \left(2 e^{S_{1} w}+3\right)^2}{216 \left(e^{S_{1} w}+1\right)^3} \,,
\end{align}
\begin{align}\label{Ap1112l}
\lim_{w\rightarrow+\infty} G_{D}^{C}&=\frac{a^4 S_{1}^2}{54} \,,\nonumber\\
\lim_{w\rightarrow-\infty} G_{D}^{C}&=0  \,.
\end{align}

\subsubsection{Energy momentum tensor}

The energy momentum tensor $T_{00}$ is given by:
\begin{multline}\label{t3}
T_{00}=\frac{a ^2 S_{1}^2}{1728}\exp \left[-\frac{a ^2}{9} \left(\frac{1}{\sqrt{e^{-S_{1} w}+1}}+2 \coth ^{-1}\left(\sqrt{e^{-S_{1} w}+1}\right)\right)\right] \times \\
\times \left[27 \left(\frac{1}{e^{S_{1} w}+1}\right)^{3/2} \left(\frac{2 \sqrt{e^{S_{1} w}}}{e^{S_{1} w}+1}+\sqrt{\sech^2\left(\frac{S_{1} w}{2}\right)}\right)-\frac{4 a ^2 e^{S_{1} w} \left(2 e^{S_{1} w}+3\right)^2}{\left(e^{S_{1} w}+1\right)^3}\right] \,.
\end{multline}

\subsubsection{Geodesic equation}

The geodesic equation is given by:
\begin{align}\label{geo3}
\ddot{w}+c_{1}^{2}\frac{\left(25 a ^2+27 \sqrt{2}\right) a ^2 S_{1}^2  e^{\frac{1}{18} a ^2 \left(\sqrt{2}+4 \coth ^{-1}\left(\sqrt{2}\right)\right)}}{5184}w\approx -c_{1}^{2} \frac{5 a ^2 S_{1} e^{\frac{1}{18} a ^2 \left(\sqrt{2}+4 \coth ^{-1}\left(\sqrt{2}\right)\right)}}{72 \sqrt{2}}.
\end{align}

\subsection{ Case $n=2$ ($\phi^{18}$)}

\subsubsection{Auxiliary function and warp factor}

The auxiliary function $W$ for this case is given by:
\begin{equation}\label{w32}
W(\phi)=\frac{S_{2}}{40}  \phi ^2 \left(5-\frac{\phi ^8}{a ^8}\right)\,,
\end{equation}
which provides the following expression for the potential $V(\phi)$:
\begin{equation}\label{V32}
V(\phi)=\frac{\left[75 \left(1-\frac{\phi ^8}{a ^8}\right)^2-2 \phi ^2 \left(\frac{\phi ^8}{a ^8}-5\right)^2\right] \left(S_{2}^2 \phi ^2\right)}{9600} \,.
\end{equation}

The warp factor $A$ takes the form:
\begin{align}\label{Ap13}
e^{2A}=\left(\frac{1-\sqrt[4]{\frac{1}{e^{S_{2} (-w)}+1}}}{\sqrt[4]{\frac{1}{e^{S_{2} (-w)}+1}}+1}\right)^{\frac{a ^2}{15}} \exp \left[-\frac{a ^2}{15}  \left(\sqrt[4]{\frac{1}{e^{S_{2} (-w)}+1}}-2 \cot ^{-1}\left(\sqrt[4]{1-\frac{1}{e^{S_{2} w}+1}}\right)\right)\right] \,.
\end{align}

\subsubsection{Ricci scalar}

The Ricci scalar, $R$, for this model is given by:
\begin{align}\label{Ap14}
R=-\frac{a^2 S_{2}^2 e^{S_{2} w} \left[a^2 \sqrt[4]{\frac{e^{S_{2} w}}{e^{S_{2} w}+1}} \left(4 e^{S_{2} w}+5\right)^2-60\right]}{720 \left(\frac{1}{e^{-S_{2} w}+1}\right)^{3/4} \left(e^{S_{2} w}+1\right)^3} \,,
\end{align}
\begin{align}\label{Ap14l}
\lim_{w\rightarrow+\infty} R&=-\frac{ a^4 S_{2}^2}{45},\nonumber\\
\lim_{w\rightarrow-\infty} R&=0,\nonumber\\
\lim_{w\rightarrow0} R&=-\frac{a^2 \left(\frac{81 a^2}{\sqrt[4]{2}}-60\right) S_{2}^2}{2880 \sqrt[4]{2}} \,.
\end{align}

\subsubsection{Einstein tensor components}

The mixed Einstein tensor components $G_{D}^{C}$:
\begin{align}\label{Ap15}
G^{\mu}_{\mu}&=\frac{a^2 S_{2}^2 e^{S_{2} w} \left[a^2 \left(4 e^{S_{2} w}+5\right)^2 \sqrt[4]{1-\frac{1}{e^{S_{2} w}+1}}-75\right]}{2400 \left(\frac{1}{e^{-S_{2} w}+1}\right)^{3/4} \left(e^{S_{2} w}+1\right)^3} \,,
\end{align}
\begin{align}\label{Ap15l}
\lim_{w\rightarrow0}G^{\mu}_{\mu}=\frac{a^2 \left(\frac{81 a^2}{\sqrt[4]{2}}-75\right) S_{2}^2}{9600 \sqrt[4]{2}} \,,
\end{align}
\begin{align}\label{Ap16}  
G^{4}_{4}=\frac{a^4 S_{2}^2 \left(4 e^{S_{2} w}+5\right)^2 \sqrt{1-\frac{1}{e^{S_{2} w}+1}}}{2400 \left(e^{S_{2} w}+1\right)^2} \,,
\end{align}
\begin{align}\label{Ap1516l}
\lim_{w\rightarrow+\infty} G_{D}^{C}&=\frac{a^4 S_{2}^2}{150},\nonumber\\
\lim_{w\rightarrow-\infty} G_{D}^{C}&=0 \,.
\end{align}

\subsubsection{Energy momentum tensor}

The energy momentum tensor $T_{00}$:
\begin{multline}\label{t4}
T_{00}=\frac{a ^2 S_{2}^2 \sqrt[4]{e^{S_{2} w}}} {9600 \left(e^{S_{2} w}+1\right)^{9/4}}\exp \left[-\frac{a ^2}{15}  \left(\sqrt[4]{\frac{1}{e^{-S_{2} w}+1}}-2 \cot ^{-1}\left(\sqrt[4]{1-\frac{1}{e^{S_{2} w}+1}}\right)\right)\right] \times
	\\
\times \left(\frac{1-\sqrt[4]{\frac{1}{e^{-S_{2} w}+1}}}{\sqrt[4]{\frac{1}{e^{-S_{2} w}+1}}+1}\right)^{\frac{a ^2}{15}} \left[75 \left(\sqrt[4]{\frac{1}{e^{S_{2} w}+1}} \sqrt[4]{e^{S_{2} w}+1}+1\right)-2 a ^2 \left(4 e^{S_{2} w}+5\right)^2 \sqrt[4]{1-\frac{1}{e^{S_{2} w}+1}}\right] \,.
\end{multline}

\subsubsection{Geodesic equation}

The geodesic equation is given by:
 \begin{multline}\label{geo4}  
\ddot{w}+c_{1}^{2}\frac{a ^2 \left(27 a ^2+25 \sqrt[4]{2}\right) \left(\frac{2+2^{3/4}}{2-2^{3/4}}\right)^{\frac{a ^2}{15}} S_{2}^2  e^{\frac{1}{30} a ^2 \left(2^{3/4}-4 \tan ^{-1}\left(\sqrt[4]{2}\right)\right)}}{9600 \sqrt{2}}w\approx -c_{1}^{2}\frac{3 a ^2 \left(\frac{2+2^{3/4}}{2-2^{3/4}}\right)^{\frac{a ^2}{15}} S_{2}e^{\frac{1}{30} a ^2 \left(2^{3/4}-4 \tan ^{-1}\left(\sqrt[4]{2}\right)\right)}}{80 \sqrt[4]{2}}.
\end{multline}

\section{Properties of Thick Brane with Power-Law Tail (PT)}\label{appendix:CC}

\subsection{ Case $n=1$ ($\phi^{10}$)}

\subsubsection{Auxiliary function and warp factor}

The auxiliary function $W$ for this case is given by:
\begin{align}\label{Ap17}
W(\phi)=-\frac{S_{1}}{96} \left[\phi  \sqrt{a ^2-\phi ^2} \left(\frac{8 \phi ^4}{a ^4}-\frac{14 \phi ^2}{a ^2}+3\right)-3 a ^2 \tan ^{-1}\left(\frac{\phi }{\sqrt{a ^2-\phi ^2}}\right)\right]\,,
\end{align}
which provides the following expression for the potential $V(\phi)$:
\begin{align}\label{Ap18}
 V(\phi)=-\frac{S_{1}^2}{27648}\left[-6 \phi  \sqrt{(a -\phi ) (a +\phi )} \left(\frac{8 \phi ^4}{a ^2}+3 a ^2-14 \phi ^2\right) \tan ^{-1}\left(\frac{\phi }{\sqrt{a ^2-\phi ^2}}\right)+9 a ^4 \tan ^{-1}\left(\frac{\phi }{\sqrt{a ^2-\phi ^2}}\right)^2\right.\nonumber\\
\left.  +\phi ^2 \left(\frac{64 \phi ^8}{a ^8}-\frac{864 \phi ^6}{a ^8}-\frac{224 \phi ^6}{a ^6}+\frac{1728 \phi ^4}{a ^6}+\frac{244 \phi ^4}{a ^4}-\frac{864 \phi ^2}{a ^4}-\frac{84 \phi ^2}{a ^2}+9\right) \left(a ^2-\phi ^2\right)\right] \,.
 \end{align}
 
 The warp factor $A$ takes the form:
\begin{align}\label{Ap19}
e^{2A}=\exp \left[\frac{a ^2}{144}  \left(-\frac{8 S_{1}}{\sqrt{S_{1}^2 w^2+64}}-3 S_{1} w \tan ^{-1}\left(\frac{1}{8} \left(\sqrt{S_{1}^2 w^2+64}+S_{1} w\right)\right)-8\right)\right] \,.
\end{align}

\subsubsection{Ricci scalar}

The Ricci scalar, $R$, for this model is given by:
\begin{align}\label{Ap20}
R&= -\frac{a^2  S_{1}^2}{5184 \left( S_{1}^2 w^2+64\right)^2 \left( S_{1}^2 w^2+ S_{1} w \sqrt{ S_{1}^2 w^2+64}+64\right)^3}\times
	\nonumber\\ 
&\times \left[45 a^2 \left( S_{1}^4 w^4+80  S_{1}^2 w^2+48  S_{1} w \sqrt{ S_{1}^2 w^2+64}+ S_{1}^3 w^3 \sqrt{ S_{1}^2 w^2+64}+1024\right)\right.\nonumber\\ 
&\left( S_{1}^2 w^2+64\right)^3 \tan ^{-1}\left(\frac{1}{8} \left(\sqrt{ S_{1}^2 w^2+64}+ S_{1} w\right)\right)^2 +120 a^2  S_{1} w \left( S_{1}^4 w^3 (3 w-2)+48  S_{1}^2 w (5 w-2) \times \right.\nonumber\\ 
&\times \left. +16  S_{1} (9 w-2) \sqrt{ S_{1}^2 w^2+64}+ S_{1}^3 w^2 (3 w-2) \sqrt{ S_{1}^2 w^2+64}+3072\right) \left( S_{1}^2 w^2+64\right)^2 \tan ^{-1}\left(\frac{1}{8} \left(\sqrt{ S_{1}^2 w^2+64} 
+ S_{1} w\right)\right)\nonumber\\ 
&+16 \left( S_{1}^8 w^6 \left(5 a^2 (2-3 w)^2-576 w\right)+16  S_{1}^6 w^4 \left(5 a^2 \left(81 w^2-84 w+20\right)-576 w (6 w+5)\right)\right.\nonumber\\ 
&+2048  S_{1}^4 w^2 \left(5 a^2 \left(27 w^2-18 w+2\right)+144 w (1-27 w)\right)+589824  S_{1}^2 w \left(\left(5 a^2-576\right) w+96\right)\nonumber\\ 
 &+ S_{1}^7 w^5 \left(5 a^2 (2-3 w)^2-576 w\right) \sqrt{ S_{1}^2 w^2+64}+48  S_{1}^5 w^3 \left(5 a^2 \left(21 w^2-20 w+4\right)-576 w (2 w+1)\right) \sqrt{ S_{1}^2 w^2+64}\nonumber\\ 
& \left. \left. +3072  S_{1}^3 w^2 \left(5 a^2 (9 w-4)+288 (1-7 w)\right) \sqrt{ S_{1}^2 w^2+64}-18874368  S_{1} (9 w-1) \sqrt{ S_{1}^2 w^2+64}-3623878656\right)\right] \,,
\end{align}
\begin{align}\label{Ap20l}
\lim_{w\rightarrow+\infty} R&=-\frac{5 \pi ^2 a^4  S_{1}^4}{2304 (  S_{1} + S_{1})^2} \,,\nonumber\\
\lim_{w\rightarrow-\infty} R&=0,\nonumber\\
\lim_{w\rightarrow0} R&=-\frac{a^2 S_{1}^2 \left[754974720 \pi ^2 a^2+16 (150994944 S_{1}-3623878656)\right]}{5566277615616} \,.
\end{align}

\subsubsection{Einstein tensor components}

The mixed Einstein tensor components $G_{D}^{C}$:
\begin{align}\label{Ap21}
G^{\mu}_{\mu}&=\frac{a^2  S_{1}^2 }{3456 \left( S_{1}^2 w^2+64\right)^2 \left[ S_{1} w \left(\sqrt{ S_{1}^2 w^2+64}+ S_{1} w\right)+64\right]^3} \times
	\nonumber\\
&\times \left\{9 a^2 \left[ S_{1} w \left(48 \sqrt{ S_{1}^2 w^2+64}+ S_{1} w \left( S_{1} w \left(\sqrt{ S_{1}^2 w^2+64}+ S_{1} w\right)+80\right)\right)+1024\right] \left( S_{1}^2 w^2+64\right)^3\right. \times
	\nonumber\\
&\times \tan ^{-1}\left(\frac{1}{8} \left(\sqrt{ S_{1}^2 w^2+64}+ S_{1} w\right)\right)^2 +24 a^2  S_{1} w \left[ S_{1}^4 w^3 (3 w-2)+48  S_{1}^2 w (5 w-2)+16  S_{1} (9 w-2) \sqrt{ S_{1}^2 w^2+64}\right.  \nonumber\\
& \left. + S_{1}^3 w^2 (3 w-2) \sqrt{ S_{1}^2 w^2+64}+3072\right] \left( S_{1}^2 w^2+64\right)^2 \tan ^{-1}\left(\frac{1}{8} \left(\sqrt{ S_{1}^2 w^2+64}+ S_{1} w\right)\right)\nonumber\\
&+16 \left( S_{1}^8 w^6 \left(a^2 (2-3 w)^2-144 w\right)+16  S_{1}^6 w^4 \left(a^2 \left(81 w^2-84 w+20\right)-144 w (6 w+5)\right)+2048  S_{1}^4 w^2 \left(a^2 (9 w (3 w-2)+2)\right.\right.\nonumber\\
&\left. +36 w (1-27 w)\right)+589824  S_{1}^2 w \left(\left(a^2-144\right) w+24\right)+ S_{1}^7 w^5 \left(a^2 (2-3 w)^2-144 w\right) \sqrt{ S_{1}^2 w^2+64}\nonumber\\
&+48  S_{1}^5 w^3 \left(a^2 (3 w-2) (7 w-2)-144 w (2 w+1)\right) \sqrt{ S_{1}^2 w^2+64}+3072  S_{1}^3 w^2 \left(a^2 (9 w-4)-504 w+72\right) \sqrt{ S_{1}^2 w^2+64}\nonumber\\
&\left.\left. -4718592  S_{1} (9 w-1) \sqrt{ S_{1}^2 w^2+64}-905969664\right)\right\} \,,
\end{align}
\begin{align}\label{Ap21l}
\lim_{w\rightarrow0}G^{\mu}_{\mu}=\frac{a^2 S_{1}^2 \left[150994944 \pi ^2 a^2+16 (37748736 S_{1}-905969664)\right]}{3710851743744} \,,
\end{align}
\begin{align}\label{Ap22}
G^{4}_{4}&=\frac{a^4  S_{1}^2}{3456 \left( S_{1}^2 w^2+64\right)^2 \left[ S_{1} w \left(\sqrt{ S_{1}^2 w^2+64}+ S_{1} w\right)+64\right]^3} \times
\nonumber\\
&\times \left\{9 \left[ S_{1} w \left(48 \sqrt{ S_{1}^2 w^2+64}+ S_{1} w \left( S_{1} w \left(\sqrt{ S_{1}^2 w^2+64}+ S_{1} w\right)+80\right)\right)+1024\right] \left( S_{1}^2 w^2+64\right)^3 \right.\nonumber\\
&\tan ^{-1}\left(\frac{1}{8} \left(\sqrt{ S_{1}^2 w^2+64}+ S_{1} w\right)\right)^2 +24  S_{1} w \left[ S_{1}^4 w^3 (3 w-2)+48  S_{1}^2 w (5 w-2)+16  S_{1} (9 w-2) \sqrt{ S_{1}^2 w^2+64}\right.\nonumber\\
&\left. + S_{1}^3 w^2 (3 w-2) \sqrt{ S_{1}^2 w^2+64}+3072\right] \left( S_{1}^2 w^2+64\right)^2 \tan ^{-1}\left(\frac{1}{8} \left(\sqrt{ S_{1}^2 w^2+64}+ S_{1} w\right)\right)\nonumber\\
& +16  S_{1}^2 w^2 \left( S_{1}^6 w^4 (2-3 w)^2+16  S_{1}^4 w^2 \left(81 w^2-84 w+20\right)+2048  S_{1}^2 (9 w (3 w-2)+2)\right.\nonumber\\
&\left.\left. +3072  S_{1} (9 w-4) \sqrt{ S_{1}^2 w^2+64}+ S_{1}^5 w^3 (2-3 w)^2 \sqrt{ S_{1}^2 w^2+64}+48  S_{1}^3 w (3 w-2) (7 w-2) \sqrt{ S_{1}^2 w^2+64}+589824\right)\right\} \,,
\end{align}
\begin{align}\label{Ap2122l}
\lim_{w\rightarrow+\infty} G_{D}^{C}&=\frac{\pi ^2 a^4 S_{1}^4}{1536 ( S_{1} +S_{1})^2} \,,\nonumber\\
\lim_{w\rightarrow-\infty} G_{D}^{C}&=0  \,.
\end{align}

\subsubsection{Energy momentum tensor}

The energy momentum tensor $T_{00}$:
\begin{align}\label{t5}
T_{00}&=\frac{a ^2 S_{1}^2}{27648} \exp \left[\frac{ a ^2}{144} \left(-\frac{8 S_{1}}{\sqrt{S_{1}^2 w^2+64}}-3 S_{1} w \tan ^{-1}\left(\frac{1}{8} \left(\sqrt{S_{1}^2 w^2+64}+S_{1} w\right)\right)-8\right)\right] \times
	\nonumber\\
&\times \left\{3 a ^2 \tan ^{-1}\left(\frac{\sqrt{\sqrt{S_{1}^2 w^2+64}+S_{1} w}}{\sqrt{\sqrt{S_{1}^2 w^2+64}-S_{1} w}}\right)\right. \times 
	\nonumber\\
&\times \left[-8 \left(3 S_{1} w \sqrt{S_{1}^2 w^2+64}+128\right) \left(\frac{1}{S_{1}^2 w^2+64}\right)^{3/2}-3 \tan ^{-1}\left(\frac{\sqrt{\sqrt{S_{1}^2 w^2+64}+S_{1} w}}{\sqrt{\sqrt{S_{1}^2 w^2+64}-S_{1} w}}\right)\right]\nonumber\\
&\left.+\frac{-262144 \left(a ^2-54\right)-48 S_{1} w \left(3 a ^2 S_{1}^3 w^3+256 \left(a ^2+18\right) \sqrt{S_{1}^2 w^2+64}+192 \left(a ^2-24\right) S_{1} w\right)}{\left(S_{1}^2 w^2+64\right)^3}\right\} \,.
\end{align}

\subsubsection{Geodesic equation}

The geodesic equation is given by:
\begin{align}\label{geo5}
\ddot{w}+c_{1}^{2}\frac{a ^2 \left(\pi  (2048+3 \pi ) a ^2+73872\right) S_1^2 e^{\frac{1}{144} a ^2 \left(S_1+8\right)}}{221184} w\approx -c_{1}^{2}\frac{(2048+3 \pi ) a ^2 S_1 e^{\frac{1}{144} a ^2 \left(S_1+8\right)}}{1152}.
\end{align}

\subsection{Case $n=2$ ($\phi^{18}$)}

\subsubsection{Auxiliary function and warp factor}

The auxiliary function $W$ for this case is given by:
\begin{align}
W(\phi)=-\frac{S_{2} \left(a ^4-\phi ^4\right)^{5/2}}{40 a ^8}\,
\end{align}
which provides the following expression for the potential $V(\phi)$:
\begin{align}\label{Ap23a}
V(\phi)=\frac{S_{2}^2 \left[\phi ^6 \left(a ^4-\phi ^4\right)^3-\frac{2}{75} \left(a ^4-\phi ^4\right)^5\right]}{128 a ^{16}} \,.
\end{align}

The warp factor $A$ takes the form:
\begin{align}\label{Ap23}
e^{2A}=\exp \left[-\frac{a ^2 \left(3 \sqrt{S_{2}^2 w^2+64}+S_{2} w\right)}{120 \sqrt{\frac{S_{2}^2 w^2}{32}+2} \sqrt{\frac{S_{2} w}{\sqrt{S_{2}^2 w^2+64}}+1}}\right] \,.
\end{align}

\subsubsection{Ricci scalar}

The Ricci scalar, $R$, for this model is given by:
\begin{multline}\label{Ap24}
R=\frac{8192 a^2 S_{2}^2 }{45 \sqrt{S_{2}^2 w^2+64} \left(\sqrt{S_{2}^2 w^2+64}+S_{2} w\right) \left(S_{2}^2 w^2+S_{2} w \sqrt{S_{2}^2 w^2+64}+64\right)^4}\times
	\\
\times \left[ 15 \sqrt{2} \sqrt{\frac{S_{2}^2 w^2+S_{2} w \sqrt{S_{2}^2 w^2+64}+64}{S_{2}^2 w^2+64}} \right. \times
\\
\times \left. \left(S_{2}^4 w^4+80 S_{2}^2 w^2+48 S_{2} w \sqrt{S_{2}^2 w^2+64}+S_{2}^3 w^3 \sqrt{S_{2}^2 w^2+64}+1024\right)-256 a^2\right] \,,
\end{multline}
\begin{align}\label{Ap24l}
\lim_{w\rightarrow+\infty} R&=0,\nonumber\\
\lim_{w\rightarrow-\infty} R&=0,\nonumber\\
\lim_{w\rightarrow0} R&=\frac{a^2 \left(15360 \sqrt{2}-256 a^2\right)S_{2}^2}{5898240} \,.
\end{align}

\subsubsection{Einstein tensor components}

The mixed Einstein tensor components $G_{D}^{C}$:
\begin{multline}\label{Ap25}
G^{\mu}_{\mu}=-\frac{1024 a^2 S_{2}^2}{75 \sqrt{S_{2}^2 w^2+64} \left(\sqrt{S_{2}^2 w^2+64}+S_{2} w\right) \left[S_{2} w \left(\sqrt{S_{2}^2 w^2+64}+S_{2} w\right)+64\right]^4} \times
	\\
\left\{75 \sqrt{\frac{2 S_{2} w}{\sqrt{S_{2}^2 w^2+64}}+2} \left[S_{2} w \left(48 \sqrt{S_{2}^2 w^2+64}+S_{2} w \left(S_{2} w \left(\sqrt{S_{2}^2 w^2+64}+S_{2} w\right)+80\right)\right)+1024\right]-1024 a^2\right\} \,,
\end{multline}
\begin{align}\label{Ap25l}
\lim_{w\rightarrow0}G^{\mu}_{\mu}=-\frac{a^2 \left(76800 \sqrt{2}-1024 a^2\right) S_{2}^2}{78643200} \,,
\end{align}
\begin{align}\label{Ap26}
G^{4}_{4}=\frac{1048576 a^4 S_{2}^2}{75 \sqrt{S_{2}^2 w^2+64} \left(\sqrt{S_{2}^2 w^2+64}+S_{2} w\right) \left[S_{2} w \left(\sqrt{S_{2}^2 w^2+64}+S_{2} w\right)+64\right]^4} \,,
\end{align}
\begin{align}\label{Ap2526l}
\lim_{w\rightarrow+\infty} G_{D}^{C}&=0,\nonumber\\
\lim_{w\rightarrow-\infty} G_{D}^{C}&=0 \,.
\end{align}

\subsubsection{Energy momentum tensor}

The energy momentum tensor $T_{00}$:
\begin{eqnarray}\label{t6}
T_{00}&=&\exp \left[-\frac{a ^2 \left(3 \sqrt{S_{2}^2 w^2+64}+S_{2} w\right)}{120 \sqrt{\frac{S_{2}^2 w^2}{32}+2} \sqrt{\frac{S_{2} w}{\sqrt{S_{2}^2 w^2+64}}+1}}\right]  
	\times \nonumber \\
&&\times \left\{\frac{a ^2 S_{2}^2}{307200} \left(1-\frac{S_{2} w}{\sqrt{S_{2}^2 w^2+64}}\right)^3 \left[75 \sqrt{2} \left(\frac{S_{2} w}{\sqrt{S_{2}^2 w^2+64}}+1\right)^{3/2}-2 a ^2 \left(\frac{S_{2} w}{\sqrt{S_{2}^2 w^2+64}}-1\right)^2\right] 
\right. \nonumber \\
&& \left.
+\frac{64 \sqrt{2} a ^2 S_{2}^2}{\left(S_{2}^2 w^2+64\right)^3 \left(\frac{S_{2} w}{\sqrt{S_{2}^2 w^2+64}}+1\right)^{3/2}}\right\} \,.
\end{eqnarray}

\subsubsection{Geodesic equation}

The geodesic equation is given by:
\begin{align}\label{geo6}
\ddot{w}+c_{1}^{2} \frac{e^{\frac{a ^2}{5 \sqrt{2}}} a ^2 \left(a ^2+75 \sqrt{2}\right) S_2^2}{230400} w\approx c_{1}^{2}\frac{e^{\frac{a ^2}{5 \sqrt{2}}} a ^2 S_2}{480 \sqrt{2}}.
\end{align}

\end{widetext}


\begin{thebibliography}{99}

\bibitem{Randall:1999ee} 
L.~Randall and R.~Sundrum,
  ``A Large mass hierarchy from a small extra dimension,''
  Phys.\ Rev.\ Lett.\  {\bf 83}, 3370 (1999).
  arXiv:hep-ph/9905221.
  
\bibitem{Randall:1999vf} 
L.~Randall and R.~Sundrum,
  ``An Alternative to compactification,''
  Phys.\ Rev.\ Lett.\  {\bf 83}, 4690 (1999).
  arXiv:hep-th/9906064.

\bibitem{Dvali:2000hr}
  G.~R.~Dvali, G.~Gabadadze and M.~Porrati,
  ``4-D gravity on a brane in 5-D Minkowski space,''
  Phys.\ Lett.\ B {\bf 485}, 208 (2000).
  arXiv:hep-th/0005016.

\bibitem{Shiromizu:1999wj}
  T.~Shiromizu, K.~i.~Maeda and M.~Sasaki,
  ``The Einstein equation on the 3-brane world,''
  Phys.\ Rev.\ D {\bf 62}, 024012 (2000).
  arXiv:gr-qc/9910076.


\bibitem{Maartens:2003tw}
  R.~Maartens,
 ``Brane world gravity,''
  Living Rev.\ Rel.\  {\bf 7}, 7 (2004).
  arXiv:gr-qc/0312059.

\bibitem{Ge2}
R.~Davies, D.~P.~George and R.~R.~Volkas,
  ``The Standard model on a domain-wall brane,''
  Phys.\ Rev.\ D {\bf 77}, 124038 (2008). 
  arXiv:hep-ph/0705.1584.

\bibitem{lin}
G.~W.~Gibbons, R.~Kallosh and A.~D.~Linde,
  ``Brane world sum rules,''
  JHEP {\bf 0101}, 022 (2001). 
  arXiv:hep-th/0011225.
  
\bibitem{PCU}
P.~Binetruy, C.~Deffayet, U.~Ellwanger and D.~Langlois,
  ``Brane cosmological evolution in a bulk with cosmological constant,''
  Phys.\ Lett.\ B {\bf 477}, 285 (2000). 
  arXiv:hep-th/9910219.

\bibitem{Ge3}
D.~P.~George and R.~R.~Volkas,
  ``Kink modes and effective four dimensional fermion and Higgs brane models,''
  Phys.\ Rev.\ D {\bf 75}, 105007 (2007). 
  arXiv:hep-ph/0612270.

\bibitem{Ge1}
D.~P.~George, M.~Trodden and R.~R.~Volkas,
  ``Extra-dimensional cosmology with domain-wall branes,''
  JHEP {\bf 0902}, 035 (2009).
  arXiv:hep-ph/0810.3746.

\bibitem{Ge5}
D.~P.~George,
  ``Stability of gravity-scalar systems for domain-wall models with a soft wall,''
  J.\ Phys.\ Conf.\ Ser.\  {\bf 259}, 012034 (2010).
  arXiv:hep-th/1010.1628.
  
    
\bibitem{de} 
O.~DeWolfe, D.~Z.~Freedman, S.~S.~Gubser and A.~Karch,
  ``Modeling the fifth-dimension with scalars and gravity,''
  Phys.\ Rev.\ D {\bf 62}, 046008 (2000).
  arXiv:hep-th/9909134.
  
  \bibitem{CG}
J.~A.~Cabrer, G.~von Gersdorff and M.~Quiros,
  ``Soft-Wall Stabilization,''
  New J.\ Phys.\  {\bf 12}, 075012 (2010).
  arXiv:hep-ph/0907.5361.

\bibitem{blg}
P.~D.~Mannheim,
  ``Brane-localized gravity,''
  Hackensack, USA: World Scientific, 337 (2005).

\bibitem{AIS}
A.~C.~Davis, S.~C.~Davis, W.~B.~Perkins and I.~R.~Vernon,
  ``Brane world phenomenology and the Z(2) symmetry,''
  Phys.\ Lett.\ B {\bf 504}, 254 (2001).
  arXiv:hep-ph/0008132.

\bibitem{DM}
D.~Yamauchi and M.~Sasaki,
  ``Brane world in arbitrary dimensions without Z(2) symmetry,''
  Prog.\ Theor.\ Phys.\  {\bf 118}, 245 (2007).
  arXiv:gr-qc/0705.2443.

\bibitem{CC}
C.~Germani and C.~F.~Sopuerta,
  ``String inspired brane world cosmology,''
  Phys.\ Rev.\ Lett.\  {\bf 88}, 231101 (2002).
  arXiv:hep-th/0202060.

\bibitem{JL}
J.~E.~Lidsey,
  ``Inflation and brane worlds,''
  Lect.\ Notes Phys.\  {\bf 646}, 357 (2004).
  arXiv:astro-ph/0305528.



\bibitem{Ge4}
D.~P.~George,
  ``Survival of scalar zero modes in warped extra dimensions,''
  Phys.\ Rev.\ D {\bf 83}, 104025 (2011).
  arXiv:hep-th/1102.0564.

\bibitem{br} 
Damien P. George, \textit{Domain-wall brane models
of an infinite extra dimension}, Ph.D. Thesis, The University of
Melbourne, Australia (2009). 

\bibitem{Rosa:2021tei}
J.~L.~Rosa, M.~A.~Marques, D.~Bazeia and F.~S.~N.~Lobo,
``Thick branes in the scalar\textendash{}tensor representation of f(R,~T) gravity,''
Eur. Phys. J. C \textbf{81}, no.11, 981 (2021). 
arXiv:2105.06101 [gr-qc].

\bibitem{Bazeia:2020jma}
D.~Bazeia, D.~A.~Ferreira, F.~S.~N.~Lobo and J.~L.~Rosa,
``Novel modified gravity braneworld configurations with a Lagrange multiplier,''
Eur. Phys. J. Plus \textbf{136},  no.3, 321 (2021).
arXiv:2011.06240 [gr-qc].

\bibitem{Rosa:2020uli}
J.~L.~Rosa, D.~A.~Ferreira, D.~Bazeia and F.~S.~N.~Lobo,
``Thick brane structures in generalized hybrid metric-Palatini gravity,''
Eur. Phys. J. C \textbf{81}, no.1, 20 (2021). 
arXiv:2010.10074 [gr-qc].

\bibitem{Rosa:2022fhl}
J.~L.~Rosa, A.~S.~Lob\~ao and D.~Bazeia,
``Impact of compactlike and asymmetric configurations of thick branes on the scalar-tensor representation of $f\left( R,T\right) $ gravity,''
Eur. Phys. J. C \textbf{82}, no.3, 191 (2022). 
arXiv:2202.10713 [gr-qc].

\bibitem{Bazeia:2022agk}
D.~Bazeia, A.~S.~Lob\~ao and J.~L.~Rosa,
``Multi-kink braneworld configurations in the scalar-tensor representation of $f(R,T)$ gravity,''
Eur. Phys. J. Plus \textbf{137}, no.9, 999 (2022). 
arXiv:2209.01928 [gr-qc].


\bibitem{Dzh}
  V.~Dzhunushaliev, V.~Folomeev and M.~Minamitsuji,
  ``Thick brane solutions,''
  Rept.\ Prog.\ Phys.\  {\bf 73}, 066901 (2010).
  arXiv:gr-qc/0904.1775.

\bibitem{Bazeia:2004dh}
  D.~Bazeia and A.~R.~Gomes,
  ``Bloch brane,''
  JHEP {\bf 0405}, 012 (2004).
  arXiv:hep-th/0403141.
  
\bibitem{deSouzaDutra:2008gm}
  A.~de Souza Dutra, A.~C.~A.~de Faria, and M.~Hott,
 ``Degenerate and critical Bloch branes,''
  Phys.\ Rev.\ D {\bf 78}, 043526 (2008).
  arXiv:hep-th/0807.0586.

\bibitem{Peyravi:2021bra}
 M.~Peyravi, N.~Riazi and F.~S.~N.~Lobo,
  ``Novel thick brane solutions with U(1) symmetry breaking,''
  Eur.\ Phys.\ J.\ C {\bf 81}, 216 (2021).
  arXiv:gr-qc/2004.05121v2.



\bibitem{81} 
Y.~B.~Zeldovich, I.~Y.~Kobzarev and L.~B.~Okun,
  ``Cosmological Consequences of the Spontaneous Breakdown of Discrete Symmetry,''
  Zh.\ Eksp.\ Teor.\ Fiz.\  {\bf 67}, 3 (1974)
  [Sov.\ Phys.\ JETP {\bf 40}, 1 (1974)].

\bibitem{86}
A.~Vilenkin,
  ``Gravitational Field of Vacuum Domain Walls and Strings,''
  Phys.\ Rev.\ D {\bf 23}, 852 (1981).

\bibitem{87}
A.~Vilenkin,
  ``Gravitational Field of Vacuum Domain Walls,''
  Phys.\ Lett.\ B {\bf 133}, 177 (1983).

\bibitem{88}
A.~Vilenkin,
  ``Cosmic Strings and Domain Walls,''
  Phys.\ Rept.\  {\bf 121}, 263 (1985).


\bibitem{Peyravi:2015bra}
  M.~Peyravi, N.~Riazi and F.~S.~N.~Lobo,
  ``Soliton models for thick branes,''
  Eur.\ Phys.\ J.\ C {\bf 76}, no. 5, 247 (2016).
  arXiv:gr-qc/1504.04603.

\bibitem{Peyravi:2015kja}
M.~Peyravi, N.~Riazi and F.~S.~N.~Lobo,
``Thick brane solitons breaking $Z_2$ symmetry,''
arXiv:1509.04577 [gr-qc].


  \bibitem{kkka} 
I.~C.~Christov, R.~J.~Decker, A.~Demirkaya, V.~A.~Gani, P.~G.~Kevrekidis, A.~Khare, A~.Saxena,
  ``Kink-kink and kink-antikink interactions with long-range tails,''
  Phys.\ Rev.\ Lett.\  {\bf 122}, 171601 (2019). 
  arXiv:hep-th/1811.07872 .

\bibitem{lrik} 
I.~C.~Christov, R.~J.~Decker, A.~Demirkaya, V.~A.~Gani, P.~G.~Kevrekidis, R.~V.~Radomskiy,
  ``Long-range interactions of kinks,''
  Phys.\ Rev.\ {\bf D 99}, 016010 (2019).
  arXiv:hep-th/1810.03590.
  

  \bibitem{kaka} 
I.~C.~Christov, R.~J.~Decker, A.~Demirkaya, V.~A.~Gani, P.~G.~Kevrekidis, A.~Khare, A~.Saxena,
  ``Kink-Antikink Collisions and Multi-Bounce Resonance Windows in Higher-Order Field Theories,''
  Commun.\ Nonlinear Sci.\ Numer.\ Simulat.\  {\bf 97}, 105748 (2021). 
  arXiv:hep-th/2005.00154v3.


  \bibitem{DeG} 
Petr A.~Blinov, Tatiana V.~Gani, Vakhid A.~Gani
  ``Deformations of Kink Tails,''
 Ann.\ Phys.\  {\bf 437}, 168739 (2022). 
  arXiv:hep-th/2008.13159v2.

\bibitem{a} D.~Bazeia, L.~Losano, and J.~M.~C.~Malbouisson,
 ``Deformed defects'', Phys. Rev. {\bf D 66}, 101701
(2002). arXiv:hep-th/0209027.

\bibitem{b} 
E.~Belendryasova and V.~A.~Gani,
``Scattering of the $\varphi^8$ kinks with power-law asymptotics,''
Commun. Nonlinear Sci. Numer. Simul. \textbf{67} (2019), 414-426.
arXiv:1708.00403 [hep-th].


\bibitem{d} N.~S.~Manton,
`` Forces between kinks and antikinks with long-range tails'', J. Phys. A: Math.
Theor. {\bf 52}, 065401 (2019). arXiv:hep-th/1810.03557.

\bibitem{e} A.~Khare and A.~Saxena,
 ``Family of potentials with power law kink tails'', J. Phys. A: Math.
Theor. {\bf 52}, 365401 (2019). arXiv:hep-th/1810.12907.

\bibitem{f} J.~G.~F.~Campos and A.~Mohammadi,
 ``Interaction between kinks and antikinks with double
long-range tails'', Phys. Lett. {\bf B 818}, 136361 (2021). arXiv:hep-th/2006.01956.

\bibitem{g} G.~P.~de Brito and A.~de Souza Dutra, 
``Multikink solutions and deformed defects'',
 Ann. Phys. {\bf 351}, 620 (2014). arXiv:hep-th/1406.1764.

\bibitem{h} D.~Bazeia, E.~Belendryasova, and V.~A.~Gani,
``Scattering of kinks of the sinh-deformed $\phi^{4}$
model'',
 Eur. Phys. J. {\bf C 78}, 340 (2018). arXiv:hep-th/1710.04993.

\bibitem{i} D.~Bazeia, L.~Losano, and G.~J.~Olmo, 
``Novel connection between lump-like structures and
quantum mechanics'', Eur. Phys. J. {\bf Plus 133}, 251 (2018). arXiv:hep-th/1806.00346.

\bibitem{AA}
J.~Molina and E.~Musaev,  ``The invariant action for solitonic 5-branes,'' 
Eur.\ Phys.\ J.\ C {\bf  82}, 978 (2022).

\bibitem{ALB}
T.~Obikhod and  I.~ Petrenko, ``The role of topological invariants in the study of the early evolution of the Universe'' ,IOSR Journal Of Applied Physics (IOSR-JAP) e-ISSN: 2278-4861.Volume 15, Issue 1 Ser. II (Jan. – Feb. 2023).

\bibitem{BM}
D.~Bazeia, R.~Menezes and R.~da Rocha,
 ``A Note on Asymmetric Thick Branes,''
  Adv.\ High Energy Phys.\  {\bf 2014}, 276729 (2014).
  arXiv: hep-th/1312.3864.
  
 
  \bibitem{gir0} 
  A. Saxena, I. C. Christov and A. Khare,
  ``Higher-Order Field Theories, and 
  Beyond," In A Dynamical Perspective on the $\phi^4$ Model: Past, Present and Future, pp. 253-279. 
  Cham: Springer International Publishing, (2019).

\bibitem{gir1} 
D. Castanede Valle and E. W. Mielke,
``Relativistic soliton collisions of axion type dark matter,'' 
Phys. Lett. {\bf B 758}, 93(2016).

\bibitem{gir3} E.~Greenwood, E.~Halstead, R.~ Poltis and D.~ Stojkovic, `` Electroweak vacua, collider phenomenology, and possible connection with dark energy,'' Phys. Rev. {\bf D 79}, No. 2, 103003 (2009).  

\bibitem{HiOr}
A.~Khare, A.~Duzgun and A.~Saxena,
``Explicit kink solutions in several one-parameter families of higher-order field theory models,''
Int. J. Mod. Phys. {\bf B 35}, no.32, 2150324(2021). 
arXiv:2103.05145 [nlin.PS].

\bibitem{gir2} 
M.~A.~Lohe,
``Soliton Structures in $P (\phi$) in Two-dimensions,''
Phys. Rev. D \textbf{20}, 3120 (1979).
  
  

\bibitem{Sa}
J.~Sadeghi and A.~Mohammadi,
  ``Shape invariance for the bent brane with two scalar fields,''
  Eur.\ Phys.\ J.\ C {\bf 49}, 859 (2007).

\bibitem{Af}
V.~I.~Afonso, D.~Bazeia and L.~Losano,
  ``First-order formalism for bent brane,''
  Phys.\ Lett.\ B {\bf 634}, 526 (2006).
  arXiv:hep-th/0601069.

\bibitem{CM}
W.~T.~Cruz, R.~V.~Maluf, L.~J.~S.~Sousa and C.~A.~S.~Almeida,
  ``Gravity localization in sine-Gordon braneworlds,''
  Annals Phys.\  {\bf 364}, 25 (2016).
  arXiv:hep-th/1412.8492.

\bibitem{DA}
D.~Bazeia and A.~R.~Gomes,
  ``Bloch brane,''
  JHEP {\bf 0405}, 012 (2004).
  arXiv:hep-th/0403141.

\bibitem{JS}
S.~Jalalzadeh and H.~R.~Sepangi,
  ``Classical and quantum dynamics of confined test particles in brane gravity,''
  Class.\ Quant.\ Grav.\  {\bf 22}, 2035 (2005).
  arXiv:gr-qc/0408004.

\bibitem{YZYX} Y.~Zhong and Y.~X.~Liu
``Pure geometric thick $f(R)$-branes: stability and localization of gravity''
Eur.\ Phys.\ J.\ {\bf C76}, 321 (2016). 
 arXiv:hep-th/1507.00630.

\bibitem{FDCR} F.~ Dahia and C.~ Romero
``Confinement and stability of the motion of test particles in thick branes,''
  Phys.\ Lett.\ {\bf B 651}, 232, (2007).
  arXiv:gr-qc/0702011.



\bibitem{gui} 
M. Guidry, Gauge Field Theories: An Introduction with applications, Wiley (1991).

\bibitem{AB}
V.~I.~Afonso, D.~Bazeia, R.~Menezes and A.~Y.~Petrov,
  ``f(R)-Brane,''
  Phys.\ Lett.\ B {\bf 658}, 71 (2007).
  arXiv:hep-th/0710.3790.
  
\bibitem{BFG}
D. Bazeia, C. Furtado and A.R. Gomes,
``Brane Structure from a Scalar Field in Warped Spacetime''
JCAP {\bf 0402}, 002 (2004).
arXiv:hep-th/0308034.



\end{thebibliography}
\end{document}